\documentclass[a4paper,11pt]{article} 
\usepackage[margin=0.75in]{geometry} 
\usepackage{amssymb}
\usepackage{amsmath,epsfig, epsf, epic, graphicx} 
\usepackage{fancybox, lscape, subfigure} 
\usepackage{bm}
\usepackage{natbib}
\usepackage{graphicx,subfigure}
\DeclareGraphicsExtensions{.eps,.pdf,.jpeg,.png}
\usepackage{caption}
\usepackage[margin=2cm]{caption}
\usepackage{setspace}
\usepackage{cleveref}
\usepackage{xcolor}
\usepackage{array}
\usepackage{wrapfig}
\usepackage{multirow}
\usepackage{tabularx}
\allowdisplaybreaks

\usepackage{floatrow}

\newcommand\independent{\protect\mathpalette{\protect\independenT}{\perp}}
\def\independenT#1#2{\mathrel{\rlap{$#1#2$}\mkern2mu{#1#2}}}

\title{Estimation and false discovery control for the analysis of environmental mixtures}
\author{Srijata Samanta and Joseph Antonelli}

\begin{document}

\maketitle

\begin{abstract}
{The analysis of environmental mixtures is of growing importance in environmental epidemiology, and one of the key goals in such analyses is to identify exposures and their interactions that are associated with adverse health outcomes. Typical approaches utilize flexible regression models combined with variable selection to identify important exposures and estimate a potentially nonlinear relationship with the outcome of interest. Despite this surge in interest, no approaches to date can identify exposures and interactions while controlling any form of error rates with respect to exposure selection. We propose two novel approaches to estimating the health effects of environmental mixtures that simultaneously 1) Estimate and provide valid inference for the overall mixture effect, and 2) identify important exposures and interactions while controlling the false discovery rate. We show that this can lead to substantial power gains to detect weak effects of environmental exposures. We apply our approaches to a study of persistent organic pollutants and find that our approach is able to identify more interactions than existing approaches. }
\end{abstract}

\section{Introduction}

Estimating the health effects of environmental mixtures is an ongoing topic of research in environmental epidemiology \citep{carlin2013unraveling, braun2016can}.  While the study of environmental mixtures brings about much promise, there are a number of difficulties associated with such studies. Mixture analyses aim to understand the overall health impact of the mixture, which exposures are associated with changes in the outcome, and whether exposures are interacting with each other. Therefore, we are faced with simultaneous challenges of performing variable selection to identify important exposures or important interactions, while also providing precise estimates of the health effect of the mixture. Both of these goals are complicated by the fact that relationships between exposures and health outcomes can be nonlinear \citep{henn2014chemical}, and that exposures can be highly correlated with one another. This has led to the development of a number of different statistical approaches that tackle some of these issues, usually based on flexible regression models \citep{davalos2017current}. We recommend \cite{lazarevic2019statistical} for a recent review of these approaches, though we highlight a number of them here. 

An early approach can be found in \cite{herring2010nonparametric}, which allowed the effect of the environmental mixture to be linear in both the main effects and interactions among all exposures. Dirichlet process prior distributions are used on the regression coefficients to identify important exposures or interactions, and to cluster coefficients, which can be helpful with highly correlated exposures. Weighted quantile sum regression \citep{carrico2015characterization} creates a score, which is a linear combination of the quantiles of each environmental exposure, and relates this to an outcome through a regression model. Each exposure is assigned a weight in this score variable, and variable selection is done by thresholding values of the weight vector. Bayesian kernel machine regression (BKMR, \cite{bobb2015bayesian}) uses Gaussian process regression to relate environmental exposures to a health outcome, and variable selection is performed on the exposures so that only the important ones contribute to the regression. This identifies important exposures in a highly flexible way, though does not identify important interactions between exposures. Other approaches have used flexible outcome regression models with explicit identification of interaction terms as one of the main goals. \cite{ferrari2020bayesian} propose a latent factor model where both the environmental exposures and outcome are a function of latent factors. This induces both main effect and interaction terms between the outcome and exposures, and shrinkage priors are used to increase shrinkage of higher order interactions. \cite{antonelli2020estimating} use semiparametric Bayesian regression models that allow for an unknown order of interactions between exposures, and provide inference on the probability that any interaction is important. Similarly, \cite{wei2020sparse} use nonlinear interaction terms that are estimated using basis function expansions and are included additively in an outcome regression model. Important exposures are chosen by evaluating the amount of variability explained by inclusion of each exposure or interaction. \cite{ferrari2019identifying} include main effect and interaction terms linearly, but allow for nonlinear relationships through the inclusion of an additional Gaussian process term. Most of these approaches are fully Bayesian, but recent proposals have used penalized likelihood estimators to identify interactions. A forward stepwise algorithm is developed in \cite{narisetty2019selection} that identifies important main effects and interactions as well as whether linear or nonlinear terms are required to model these effects. Group lasso was adapted for environmental mixtures in a way that enforces strong heredity of interactions and allows for nonlinear relationships in \cite{boss2020hierarchical}. Both approaches identify important main effects and interactions, however, they do not provide inference on the resulting effect estimates.

None of the existing approaches explicitly control the false discovery rate (FDR). Given that environmental mixtures have potentially weak signals with outcomes of interest, it is important to control the FDR in order to maximize power to detect small effects while avoiding false positives. There has been a spike in interest in controlling FDR in high-dimensional regression problems \citep{g2013sequential, barber2015controlling, huang2017controlling}, which have shown that FDR can be controlled when using high-dimensional models such as the lasso \citep{tibshirani1996regression}. Other recent advancements have focused on performing inference for regression coefficients in high-dimensional models. Post-selection inference on coefficients from lasso models was first developed in \cite{lockhart2014significance}. These approaches focus on conditional inference, which aims to provide inference conditional on a chosen model, and does not attempt to perform inference on the entire vector of unknown regression coefficients. A recent approach, termed the debiased lasso \citep{van2014asymptotically}, instead constructs a de-sparsified lasso estimator that has a Gaussian limiting distribution for the entire vector of regression coefficients, irrespective of any chosen model.  

In this manuscript, we develop new approaches that address the aforementioned issues of existing estimators for environmental mixtures. By utilizing recently developed approaches in the high-dimensional statistics literature, we are able to simultaneously provide inference on the effect of the mixture on health outcomes and identify important exposures and interactions with theoretical guarantees on the false discovery rates. We show that this can lead to substantial gains in power to detect moderate to weak signals in the data, without sacrificing in terms of estimation accuracy or inference on the resulting mixture effect. 

\section{Notation and statement of problem}

Throughout, we observe $n$ i.i.d samples of $(Y, \boldsymbol{X}, \boldsymbol{C})$. Here $Y$ is the response variable, $\boldsymbol{X}$ is a $p-$dimensional vector of environmental exposures and $\boldsymbol{C}$ is a $q-$dimensional vector of covariates. We target the effect of environmental exposures conditional on $\boldsymbol{C}$, and our goals are three-fold: 1) to estimate the effect of the exposures on the outcome, 2) perform valid inference on the resulting mixture effect, and ) perform feature selection and control FDR. In traditional model selection or testing applications, the false discovery rate is defined as 
$\text{FDR} = E \bigg( \frac{\text{number of false discoveries}}{\text{number of discoveries}} \bigg),$
with the convention being that $0/0 = 0$. Our interest is the identification of important environmental exposures, and whether there are interactions among the environmental exposures. We define a pair of environmental exposures $(X_j, X_k)$ to have an interaction if for some $(x_j, x_k)$ and $(x_j', x_k')$:
\begin{align}
& E(Y \vert \boldsymbol{X}_{-jk} = \boldsymbol{x}_{-jk}, X_j=x_j, X_k=x_k, \boldsymbol{C}) - E(Y \vert \boldsymbol{X}_{-jk} = \boldsymbol{x}_{-jk}, X_j=x_j', X_k=x_k, \boldsymbol{C}) \nonumber \\
\neq & E(Y \vert \boldsymbol{X}_{-jk} = \boldsymbol{x}_{-jk}, X_j=x_j, X_k=x_k', \boldsymbol{C}) - E(Y \vert \boldsymbol{X}_{-jk} = \boldsymbol{x}_{-jk}, X_j=x_j', X_k=x_k', \boldsymbol{C}). \label{eqn:interactionDefinition}
\end{align}

\noindent Note here that $\boldsymbol{X}_{-jk}$ represents the $p-2$ remaining environmental exposures. The inequality \ref{eqn:interactionDefinition} states that the effect of changing $X_j$ from $x_j$ to $x_j'$ depends on the value of $X_k$. To identify important exposures and interactions among the exposures, we utilize the following model for the outcome:
\begin{align}
    g(E(Y \vert \boldsymbol{X}, \boldsymbol{C})) = \beta_0 + \boldsymbol{C} \boldsymbol{\beta}_C + \sum_{j=1}^p f(X_j) + \sum_{j_1 = 2}^p \sum_{j_2 < j_1} f(X_{j_1}, X_{j_2}). \nonumber
\end{align}
This model has been shown to work well at modeling the effect of environmental mixtures \citep{wei2020sparse}. We utilize basis function expansions for each of these functions to allow for nonlinear effects of the exposures on the outcome. %Any basis functions could apply here, though we restrict to natural splines throughout the rest of this paper. We model each main effect function with $k$ degrees of freedom splines, and the interaction functions are modeled using the $k^2$ basis functions represented by tensor products of the basis functions used for the main effects.
Any basis functions could apply here, though we restrict to polynomial basis functions throughout the rest of this paper. We model each main effect function using polynomials with $k$ degrees of freedom, and the interaction functions are modeled using the $k^2$ basis functions represented by tensor products of the basis functions used for the main effects. We perform an orthogonal projection of the interaction basis functions onto the space spanned by the basis functions of the two corresponding main effect functions so that they only capture information that can not be captured by the main effect functions. If we let $\boldsymbol{Y}$ be the $n$-dimensional vector of outcomes $Y_i$ for $i=1, \dots, n$, then we can write
\begin{align}
    g(E(\boldsymbol{Y} \vert \boldsymbol{X}, \boldsymbol{C})) &= \beta_0 + \boldsymbol{C} \boldsymbol{\beta}_C + \sum_{j=1}^p \boldsymbol{X}_j^* \boldsymbol{\beta}_j + \sum_{j_1 = 2}^p \sum_{j_2 < j_1} \boldsymbol{X}_{j_1,j_2}^* \boldsymbol{\beta}_{j_1, j_2} \nonumber \\
    &= \beta_0 + \boldsymbol{C} \boldsymbol{\beta}_C + \boldsymbol{X}_{main}^* \boldsymbol{\beta}_{main} +  \boldsymbol{X}_{int}^* \boldsymbol{\beta}_{int}, \label{eqn:MainModel}
\end{align}
where $\boldsymbol{X}_j^*$ is an $n \times k$ matrix of basis functions evaluated at the observed data for $X_j$, and $\boldsymbol{X}_{j_1,j_2}^*$ is the corresponding $n \times k^2$ matrix for the interaction function of $X_{j_1}$ and $X_{j_2}$. We let the total number of parameters in our model be defined by $p_t$. 

We focus on two distinct forms of false discovery rates. Let $\mathcal{S}$ represent the set of indices in $\{1, \dots p \}$ of $\boldsymbol{X}$ that have any conditional (on $\boldsymbol{C}$) association with $Y$. Formally, $\mathcal{S}$ is the set of indices such that $Y \independent \boldsymbol{X}_{-\mathcal{S}} \vert \boldsymbol{X}_{\mathcal{S}}, \boldsymbol{C}$. Further, let $\mathcal{S}_{int}$ represent the set of all pairs of indices that have an interactive effect on $\boldsymbol{Y}$ as defined in (\ref{eqn:interactionDefinition}). Let $\widehat{\mathcal{S}}$ and $\widehat{\mathcal{S}}_{int}$ be the estimates of $\mathcal{S}$ and $\mathcal{S}_{int}$, respectively. Two quantities that we target are given by
\begin{align}
    \text{FDR} &= E \bigg( \frac{\# \{j: j \in \widehat{\mathcal{S}} \backslash \mathcal{S}\}}{\# \{j: j \in \widehat{\mathcal{S}}\}} \bigg) \label{eqn:FDRmain} \\
     \text{FDR}_{int} &= E \bigg( \frac{\# \{j_1, j_2: (j_1, j_2) \in \widehat{\mathcal{S}}_{int} \backslash \mathcal{S}_{int}\}}{\# \{j_1, j_2: (j_1, j_2) \in \widehat{\mathcal{S}}_{int}\}} \bigg). \label{eqn:FDRint}
\end{align}
The first of these two quantities is the traditional FDR, while the second of these two quantities is an interaction specific false discovery rate that we adopt for the problem of environmental mixtures. Our main focus will be to control $\text{FDR}_{int}$ as interaction detection is one of the main goals of our approach. We discuss the implications of controlling $\text{FDR}_{int}$ on FDR in Section \ref{sec:implications}. 

\section{Approaches to controlling FDR with environmental mixtures}

In this section, we discuss two approaches to controlling the false discovery rate for interactions using (\ref{eqn:MainModel}). We exploit two recent advancements in the high-dimensional statistics literature to determine the set of interactions that are important in our data. 

\subsection{Using debiased lasso}
\label{debiased}

This method involves constructing confidence intervals for the effect of each exposure and subsequently performing variable selection with focus on the situation where $p_t >> n$. If we are strictly interested in estimating the coefficients in (\ref{eqn:MainModel}), the group lasso is a natural option as the $k$ basis functions for each $X_j$ in $\boldsymbol{X}_j ^*$ have a group structure. The corresponding estimator enforces that all of the coefficients in any group are simultaneously zero or non-zero.  For instance, if using polynomials as basis functions, it would be natural to assume that all the components of a group $\boldsymbol{X}_j^*$ or those of  an interaction group $\boldsymbol{X}_{j_1 , j_2}^*$ are important (or not) simultaneously. In the environmental mixtures problem there will be $p + \binom{p}{2}$ groups in total,where the first $p$ groups are of size $k$ and the remaining $\binom{p}{2}$ groups are of size $k^2$. If $p_t < n$, one can easily find confidence intervals for the entire vector of coefficients $\boldsymbol{\beta} = (\boldsymbol{\beta}_C, \boldsymbol{\beta}_{main}, \boldsymbol{\beta}_{int})$ using ordinary least squares theory. However, with $p_t >> n$ this is no longer possible, and therefore we use the debiased version of the lasso estimator, $\widehat{\beta}_d$ proposed and studied in recent works \citep{van2014asymptotically}.

Let $\boldsymbol{D}$ be the design matrix that contains all columns in $\boldsymbol{X}_{main}^*, \boldsymbol{X}_{int}^*$ and $\boldsymbol{C}$.  
% For the model,
% \begin{align}
%     \boldsymbol{Y}  &= \beta_0 + \boldsymbol{C} \boldsymbol{\beta}_C + \sum_{j=1}^p \boldsymbol{X}_j^* \boldsymbol{\beta}_j + \sum_{j_1 = 2}^p \sum_{j_2 < j_1} \boldsymbol{X}_{j_1,j_2}^* \boldsymbol{\beta}_{j_1, j_2} +\boldsymbol{\varepsilon}, \label{eqn2:MainModel}
% \end{align} 
% where $\varepsilon \sim N(0, \sigma^2 I_p)$ 
For the model $\boldsymbol{Y} = \boldsymbol{D}\boldsymbol{\beta} + \varepsilon$ we consider the debiased lasso estimator given by
\begin{align*}
    \widehat{\boldsymbol{\beta}}_d = \widehat{\boldsymbol{\beta}}_{\lambda} + \frac{1}{n} \boldsymbol{\Theta} \boldsymbol{D}^T(\boldsymbol{Y}-\boldsymbol{D}\widehat{\boldsymbol{\beta}}_{\lambda}), %= \boldsymbol{\beta} + \frac{1}{n}\Theta \boldsymbol{D}^T\varepsilon + (I- \frac{1}{n}\Theta \boldsymbol{D}^T\boldsymbol{D})(\widehat{\boldsymbol{\beta}}_{\lambda}-\boldsymbol{\beta})
\end{align*}
where $\widehat{\boldsymbol{\beta}}_{\lambda}$ is the group lasso estimate and $\boldsymbol{\Theta}$ is an approximate inverse of $\widehat{\boldsymbol{\Sigma}} = \frac{1}{n}\boldsymbol{D}^T\boldsymbol{D}$. We estimate $\boldsymbol{\Theta}$ using an approach considered in \cite{van2014asymptotically}, which they show leads to $\widehat{\boldsymbol{\beta}}_d \sim \mathcal{N}( \boldsymbol{\beta}, \frac{\sigma^2}{n}\Theta \widehat{\boldsymbol{\Sigma}} \boldsymbol{\Theta}^T) $. Using a similar approach to \cite{javanmard2019false}, we will leverage this distributional result to control $\text{FDR}_{int}$. Specifically, for each $(j_1,j_2)$ combination, let us consider testing $H_{0,j_1j_2}: \boldsymbol{\beta}_{j_1 j_2} = 0 \quad
    \text{vs}\quad H_{1,j_1j_2} : \boldsymbol{\beta}_{j_1j_2} \neq 0 $. The test statistic for the $(j_1,j_2)$ interaction, $T_{j_1 j_2}$ is defined as
$T_{j_1 j_2} = \widehat{\boldsymbol{\beta}}_d ^{(j_1 j_2)'} \left(\boldsymbol{\Omega}_{(j_1 j_2)}\right)^{-1} \widehat{\boldsymbol{\beta}}_d ^{(j_1 j_2)} \overset{H_0}{\sim}
  \chi_{k^2}^2$ where $\widehat{\boldsymbol{\beta}}_d ^{(j_1 j_2)}$ represents the coefficient estimates corresponding to the $(j_1,j_2)$ interaction and  $\boldsymbol{\Omega}_{(j_1 j_2)}$ represents the covariance matrix of $\widehat{\boldsymbol{\beta}}_d ^{(j_1 j_2)}$. We reject $H_{0,j_1 j_2}$ if for a given cut-off value $t_0$, $T_{j_1 j_2} > t_0$. Let $R(t)$ be the number of rejections for a given threshold $t$, i.e. $R(t) := \sum_{j_1 < j_2} I(T_{j_1 j_2} >t)$. To control $\text{FDR}_{int}$ at a pre-assigned level $q \in [0,1]$, we use a cut-off value $t_0$ defined as
$$t_0 = \inf\left\{t>0 :  \frac{\binom{p}{2}(1-C (t))}{\text{max}(1,R(t))} \leq q\right\}$$
where $C(\cdot)$ is the CDF of the chi-squared distribution with $k^2$ degrees of freedom. The quantity  being controlled in the definition of $t_0$ is the expected number of rejections under the global null hypothesis that no interactions are important divided by the total number of rejections. This is an extension of the false discovery controlling procedure developed in \cite{javanmard2019false} to the group setting. While our primary focus is on $\text{FDR}_{int}$, we are also interested in traditional FDR control and can therefore perform an analogous procedure on the main effect groups for each exposure, denoted by $\boldsymbol{\beta}_{j}$. In Section \ref{sec:sim} we show empirically that both $\text{FDR}_{int}$ and FDR are controlled under different scenarios using this approach.
 
%  Using similar ideas one can also considering testing $H_{0,j} : \boldsymbol{\beta}_{j} = 0\;
%     \text{vs}\; H_{1,j} : \boldsymbol{\beta}_{j} \neq 0$ to test for important main effects.
% The test static for the $j-$th main effect group will then be defined as 
% $T_{j} = \widehat{\boldsymbol{\beta}}_d ^{j'} \left(\boldsymbol{\Omega}_{j}\right)^{-1} \widehat{\boldsymbol{\beta}}_d ^{j} \overset{H_0}{\sim}
%   \chi_{k}^2$ and the cut-off value, $t_0^{'}$ for rejection of  $H_{0,j}$ will be defined by 
%   $$t_0^{'} = \inf\left\{t>0 :  \frac{p(1-C^{'} (t))}{\text{max}(1,R^{'}(t))} \leq q\right\}$$
% where $C^{'}(t)$ is the CDF of the chi-squared distribution with $k$ degrees of freedom and $R^{'}(t) := \sum_{j=1}^p I(T_j > t)$. 

\subsection{Using data splitting and knockoffs}
\label{knockoff}
A different approach entails using a model selection procedure first, then performing inference conditional on this chosen model. One can not simply ignore the fact that the model was chosen adaptively when subsequently performing inference \citep{leeb2005model}. The easiest way to bypass this issue is to use data splitting \citep{cox1975note}, which involves splitting the data into two separate parts: one that is used to select a model, and one to perform inference on the chosen model. Assuming that the procedure used for inference in the second stage is valid, data splitting leads to conditionally valid inference, sometimes referred to as selective inference. This differs from the previous approach that targets marginal inference, and we will highlight differences between these two approaches in more detail in Section \ref{sec:tradeoff}. 

For model selection, we adapt the knockoffs procedure \citep{barber2015controlling} to allow for grouped variable selection and control of false discovery rates. This procedure gives finite-sample guarantees of false discovery control without assuming correct specification of $Y \vert \boldsymbol{X}, \boldsymbol{C}$. The knockoffs procedure requires knowledge of the distribution of $\boldsymbol{X}$, but has been shown to work well even if this distribution is estimated \citep{barber2018robust}, as is typically the case. Throughout the rest of this section, when referring to model selection using the knockoffs procedure it is implied that we are restricting attention to the $n_1$ subjects used in the first stage model selection procedure. The remaining $n-n_1$ subjects will be used for inference in the second stage. The knockoffs procedure aims to construct a ``knockoff'' version of the original variables, which we will denote by $\widetilde{\boldsymbol{D}} = [\widetilde{\boldsymbol{X}}_{main}^*, \widetilde{\boldsymbol{X}}_{int}^*, \widetilde{\boldsymbol{C}}]$. The first condition required on the knockoff variables is that given $\boldsymbol{D}$, $Y$ is independent of $\widetilde{\boldsymbol{D}}$.
%$Y \independent \widetilde{\boldsymbol{D}} \vert \boldsymbol{D}$. 
This implies that the knockoff versions of the variables do not have any association with the outcome once we have conditioned on the original versions of the variables. Knockoffs can be marginally associated with the outcome, but this association is only through the association with any of the original variables that are associated with the outcome. A second condition on the knockoffs is that for any $j$, the pair of random variables $(D_j, \widetilde{D}_j)$ are exchangeable conditional on the other variables and their knockoff counterparts. We utilize the \texttt{knockoff} R package to construct knockoff variables that satisfy both properties, and refer readers to \cite{EC-ea:2018} for technical details. 

To use the knockoffs procedure, we estimate a modified version of model (\ref{eqn:MainModel}), where we include both the original and knockoff versions of the variables:
\begin{align*}
    g(E(\boldsymbol{Y} \vert \boldsymbol{X}, \boldsymbol{C})) &= \beta_0 + \boldsymbol{D} \boldsymbol{\beta} + \widetilde{\boldsymbol{D}} \widetilde{\boldsymbol{\beta}}
    % &= \beta_0 + \boldsymbol{C} \boldsymbol{\beta}_C + \sum_{j=1}^p \boldsymbol{X}_j^* \boldsymbol{\beta}_j + \sum_{j_1 = 2}^p \sum_{j_2 < j_1} \boldsymbol{X}_{j_1,j_2}^* \boldsymbol{\beta}_{j_1, j_2} \\
    % & + \widetilde{\boldsymbol{C}} \widetilde{\boldsymbol{\beta}}_C + \sum_{j=1}^p \widetilde{\boldsymbol{X}}_j^* \widetilde{\boldsymbol{\beta}}_j + \sum_{j_1 = 2}^p \sum_{j_2 < j_1} \widetilde{\boldsymbol{X}}_{j_1,j_2}^* \widetilde{\boldsymbol{\beta}}_{j_1, j_2}
\end{align*}
We obtain estimates of the parameters in this model using group lasso, which we denote by $\widehat{\boldsymbol{\beta}}$ and $\widehat{\widetilde{\boldsymbol{\beta}}}$, though any estimate of these parameters would apply. The group lasso penalty zeroes out groups of coefficients together, and we will use the parameters corresponding to each main effect and each interaction effect as the groups of coefficients in our model. For the knockoffs procedure, we need to construct statistics $W_{j_1, j_2}$ corresponding to the interaction effects that satisfy two properties. For any interaction effect that does not have an association with the outcome, we need $W_{j_1, j_2}$ to be symmetric about zero. If the interaction has an association with the outcome, we want $W_{j_1, j_2}$ to be large and positive. The most natural choice is to set $W_{j_1, j_2} = ||\widehat{\boldsymbol{\beta}}_{j_1, j_2}||_2 - ||\widehat{\widetilde{\boldsymbol{\beta}}}_{j_1, j_2}||_2$.
Clearly, if there is a significant effect of the interaction for exposures $j_1$ and $j_2$, then we expect $\widehat{\boldsymbol{\beta}}_{j_1, j_2}$ to have larger absolute values than $\widehat{\widetilde{\boldsymbol{\beta}}}_{j_1, j_2}$. If there is no effect of interaction $(j_1, j_2)$ then we expect these two quantities to behave similarly as both the original and knockoff counterpart have the same relationships with all remaining variables. Once these statistics are defined, we can proceed with selecting important interactions. We can find a threshold according to the following:
\begin{align*}
    \tau = \text{inf} \bigg\{t>0: \frac{\# \{ (j_1, j_2): W_{j_1, j_2} \leq -t \}}{\# \{(j_1, j_2): W_{j_1, j_2} \geq t \}} \leq q \bigg\},
\end{align*}
where $q$ is the target FDR rate. The selected interactions are then the set of all $(j_1, j_2)$ such that $W_{j_1, j_2} \geq \tau$. This quantity has been shown to control a modified, and less stringent form of false discoveries that adds $1/q$ in the denominator of \ref{eqn:FDRint}. If explicit control of $\text{FDR}_{int}$ is required, then $\tau$ can be replaced with
\begin{align*}
    \tau_f = \text{inf} \bigg\{t>0: \frac{1 + \# \{ (j_1, j_2): W_{j_1, j_2} \leq -t \}}{\# \{(j_1, j_2): W_{j_1, j_2} \geq t \}} \leq q \bigg\}.
\end{align*}
Empirically we have seen that using $\tau_f$ can be overly conservative, and therefore we proceed with $\tau$ that controls the modified false discovery rate. Performing this procedure will give us a list of significant interactions while controlling FDR$_{int}$ to be $q$. 

Note that our primary focus is the selection of interactions and we have restricted attention to controlling FDR$_{int}$. We will be performing inference on the chosen model in the second stage of the data splitting procedure, which requires us to select important exposures from the main effect component of the model as well. We will perform an analogous procedure to select the important main effect parameters using the knockoffs methodology, which will leave us with a reduced model that we can estimate in the second stage using traditional regression techniques. 

\subsection{Trade-offs of the two approaches}
\label{sec:tradeoff}
One advantage of the knockoffs procedure is that it is exact and does not rely on asymptotic approximations. Additionally, it does not depend on assumptions on the conditional distribution of  $Y \vert \boldsymbol{X}, \boldsymbol{C}$ making it robust to model misspecification \citep{EC-ea:2018}. The main assumption of the knockoffs procedure is correct specification of the distribution of $(\boldsymbol{X}, \boldsymbol{C})$. This may seem like a strong assumption, however, recent work has shown that knockoffs are somewhat robust to this assumption \citep{barber2018robust}. One issue with the knockoffs procedure is that it does not perform inference on the chosen exposures. We are combining the knockoffs procedure with data splitting to circumvent this issue, however, this can lead to a loss of power since we will be performing model selection on a subset of the data. The debiased lasso does not suffer from this issue as it uses the full sample for identifying the important exposures and interactions. This approach, however, makes a number of strong assumptions. First, we are relying on asymptotic approximations for the debiased lasso estimate, and its performance can deteriorate in small sample sizes. Additionally, the debiased lasso assumes the number of important variables in the model is of the order $o(\sqrt{n}/\log(p))$ which is more restrictive than what the regular lasso requires. The knockoffs procedure makes no such assumptions about the degree of sparsity in the model or about the magnitude of the nonzero coefficients. Overall, the knockoffs procedure relies on weaker assumptions than the debiased lasso, however, it comes at the cost of sample splitting in order to perform inference. 

The two approaches also differ in the type of inference that these are able to perform. As with all data splitting procedures, our inferential procedure is conditionally valid. That implies that, conditional on our chosen model, our confidence intervals will contain the true parameters at roughly the nominal rate. This differs from marginal inference, which aims to provide confidence intervals that have nominal coverage rates across all data sets, not conditional on any chosen model. The debiased lasso approach provides inference on the estimated exposure effects without conditioning on a chosen model, and is targeting marginal coverage rates. In the following section, we will evaluate the ability of each approach to provide marginally valid confidence intervals, however, it is important to clarify that the data splitting approaches are not guaranteed to achieve nominal interval coverages. 

\subsection{Implications for traditional FDR control}
\label{sec:implications}

Our approaches focus on controlling FDR$_{int}$, but it is of interest to see what implications this has for traditional FDR as defined in (\ref{eqn:FDRmain}). The first issue to understand is that our selected interactions imply a set of important exposures. For instance, if we select exposure interactions (1,2) and (2,3), then exposures 1, 2 and 3 are deemed important for traditional FDR. The second important factor to consider is that even though our focus is on interaction selection, we are additionally choosing important exposures from the main effects component of our model, which could additionally affect FDR control.

First we consider the implications of the selected interactions on FDR control. For this discussion, we ignore any exposures that are selected from the main effects component of the model and only restrict attention to the selected interactions. 
We define the false discovery proportions for any particular data set as $\text{FDP} = \frac{\#\{ j: j \in \widehat{\mathcal{S}} \backslash \mathcal{S}\}}{\# \{j: j \in \widehat{\mathcal{S}}\}}$ and $\text{FDP}_{int} = \frac{\# \{j_1, j_2: (j_1, j_2) \in \widehat{\mathcal{S}}_{int} \backslash \mathcal{S}_{int}\}}{\#\{ j_1, j_2: (j_1, j_2) \in \widehat{\mathcal{S}}_{int}\}}$, noting that FDR = $E(\text{FDP})$ and FDR$_{int}$ = $E(\text{FDP}_{int})$. Suppose that all possible interactions between exposures 1 through 4 are significant, and that these are the only important effects of the exposures. This leads to $6$ true interactions in $\mathcal{S}_{int}$ and $4$ true effects in $\mathcal{S}$. Suppose further that we correctly identify the 6 important interactions, but also include the interaction between exposures 5 and 6. This leads to $\text{FDP}_{int} = 1/7$, and $\text{FDP} = 2/6$, showing that $\text{FDP}$ can exceed $\text{FDP}_{int}$. Despite this, we can bound $\text{FDP}$ above by a function of $\text{FDP}_{int}$. First, we can write $\text{FDP}_{int}$ as
\begin{align*}
    \text{FDP}_{int} &= \frac{\text{\# false interaction discoveries}}{\text{\# interaction discoveries}} = \frac{\frac{1}{2}\text{(\# false discoveries)} + O_n}{\frac{1}{2}\text{(\# discoveries)} + O_d},
\end{align*}
where $O_n$ and $O_d$ are positive numbers accounting for the difference between the number of interaction discoveries and $1/2$ times the number of overall discoveries implied by the interactions. Selecting interactions (1,2) and (2,3) will give 2 important interactions, 3 important exposures overall, and $O_d = 1/2$. We show in Appendix B that this implies
$\text{FDP} \leq \text{FDP}_{int} \Big( 1 + \frac{2 O_d}{\text{\# discoveries}} \Big).$ This shows that FDP can be larger than $\text{FDP}_{int}$,
but only by an inflation factor that is a function of $O_d$ and the overall number of discoveries implied by our interactions. Both of these are known quantities, so the user can know how much worse FDP can be relative to $\text{FDP}_{int}$. A simple application of the Cauchy-Schwartz inequality can then be used to show that
$$E(FDP) \leq \sqrt{E(FDP_{int}^2) E \Bigg( \Big[ 1 + \frac{2 O_d}{\text{\# discoveries}} \Big]^2 \Bigg)}.$$
This shows that FDR is controlled by a quantity that is a function of both the interaction FDP and $O_d$, though this is not guaranteed to be below a pre-specified threshold of $q$. 

While the previous result provides intuition for the impact that controlling FDR$_{int}$ has on FDR control, it ignores the role of main effect parameters in our model. We can also perform variable selection on the main effect parameters, leaving us two avenues by which an exposure can be selected: through inclusion in an interaction, or through the main effect parameters. Unfortunately, even if both of these selection procedures are done in a way that respectively controls FDR, it does not guarantee that the union of the two sets of selected exposures controls FDR. If control of FDR is required, then we recommend procedures that specifically target FDR, and do not attempt to control FDR$_{int}$ simultaneously. Despite this, we will evaluate the extent to which our procedures control FDR even though our primary target of inference is FDR$_{int}$. We see empirically in Section \ref{sec:sim} that even though we can not theoretically control FDR due to the reasons described above, our approaches generally control FDR at the desired rate of $q$.

\section{Simulation study}
\label{sec:sim}

In this section we evaluate the performance of the two proposed methods under different simulation scenarios. We use two versions of our knockoffs estimator, denoted by K-Full and K-Split. K-Full uses the full data for both exposure/interaction selection and inference, while K-Split uses data splitting. K-Full is less theoretically justified, but should have larger power to detect important effects. For both knockoffs procedures, we use the threshold value $\tau$ that controls the modified false discovery rate, which is slightly less stringent than those defined in (\ref{eqn:FDRmain}) and (\ref{eqn:FDRint}). We also evaluate alternative approaches prevalent in the literature, namely BKMR (\cite{bobb2015bayesian}), NLinteraction (\cite{antonelli2020estimating}) and HiGlasso (\cite{boss2020hierarchical}). BKMR is a widely used Bayesian approach that uses Gaussian processes to flexibly model the effect of the mixture, while simultaneously identifying important mixture components. NLinteraction is also a flexible Bayesian approach that additionally performs interaction selection. HiGlasso is a penalized likelihood approach identifies important exposures and interaction pairs. We consider a range of data generating mechanisms, sample sizes, and exposure dimension listed in Table \ref{truefuncs}. Scenario 1 is a sparse situation with nonlinear interactions between a small number of exposures, scenario 2 has only linear main effects effects and no interactions, and scenario 3 is a less sparse setting where nearly every exposure has an effect on the outcome. 

\begin{table}[htbp]
    \centering
    \caption{List of simulation scenarios considered.}
    % \rowcolors{5}{}{gray!10}
    \resizebox{1.0\textwidth}{!}{%
    \begin{tabular}{*{4}{c}}
        \hline
         & $n$ & $p$ & True function, $f(\cdot)$\\
        \hline
       Scenario 1  & (200,500,1000) & (10,20) & $0.3x_7 + 0.5(x_8^2)  - 0.2x_5 + 0.4x_5^2 + 2.5x_1x_4^2 + 1.8x_2x_3$ \\
       Scenario 2  & (200,500,1000) & (10,20) & $0.5x_1 + 0.4x_2 + 0.3x_3 + 0.2x_2 + 0.1x_5$\\
      %Scenario 3 & (200,500,1000) & (10,20) & $2.3x_1 + 1.9x_1^2 - 0.2x_2 + %0.4x_2^2  - 1.5x_4 - 1.5x_4^2 + 0.05x_5 + 1.2x_7 + 0.8x_8 + x_9 + %0.2x_{10} + 0.1x_{10}^2$\\
      Scenario 3 & (200,500,1000) & (10,20) & $0.3\exp{(x_1)} + 0.5\sin(0.7x_2) - 0.2x_4 + 0.07x_5 + 0.4x_6^2 + 0.6x_8 - 0.3x_9$\\   
        \hline
    \end{tabular}
     }%
     \label{truefuncs}
\end{table}

We generate exposures from a $\mathcal{N}_p(\textbf{0},\boldsymbol{\Sigma}_X)$ distribution such that $\{\boldsymbol{\Sigma}_X\}_{jk}$ is set to $0.5^{|j-k|}$, and generate $C$ from a standard normal distribution. The response (outcome) is defined as $ Y = f(X_1,X_2,\ldots,X_p) + C\beta_C + \epsilon,$ where $\epsilon \sim N(0,I).$ 
We evaluate FDR$_{int}$ and power for interactions, FDR and power to detect exposures with any association with the outcome, the mean squared error (MSE) for estimating $f(\boldsymbol{X})$, and 95\% interval coverage probabilities for $f(\boldsymbol{X})$. We set $q=0.2$ for all approaches that control FDR. We restrict attention to results for $p=10$ exposures only. Results for $p=20$ and for other data generating mechanisms can be found in Appendix A, though the overall conclusions remain the same.

\subsection{Identifying important exposures and interactions}

\begin{figure}[htbp]
    \centering
    \includegraphics[width = 0.75\linewidth]{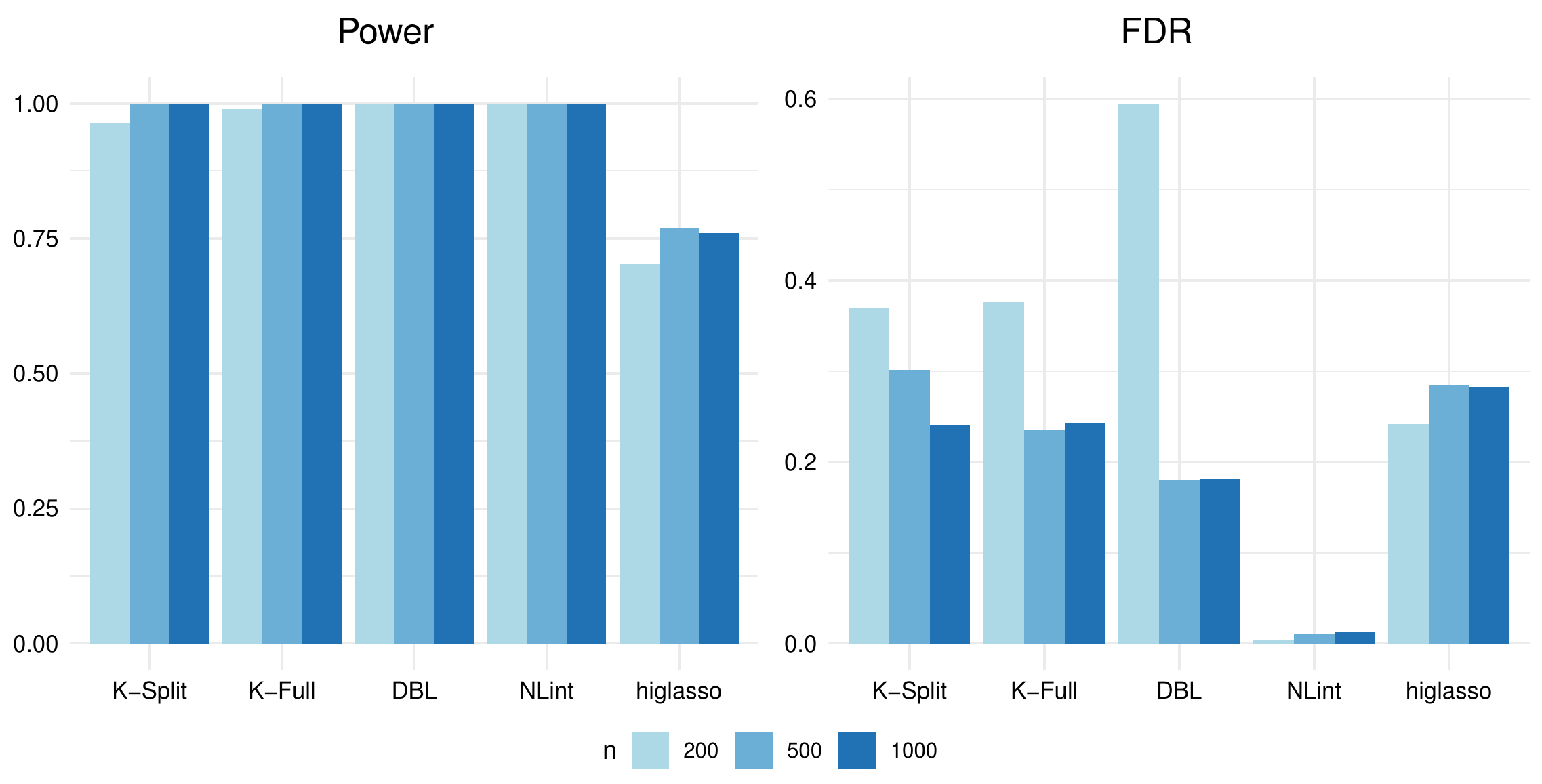}
    \caption{Empirical Power and FDR for interactions for simulation scenario 1. The left panel shows the power to detect interactions, while the right panel shows FDR$_{int}$}
    \label{fig:int}
\end{figure}

Figure \ref{fig:int} shows FDR$_{int}$ and the power to detect important interaction pairs from simulation scenario 1. Note that BKMR is not included in these results because it does not explicitly identify important interactions, and only provides visual tools to assess plausible interactions. We see that DBL does a very good job at controlling $\text{FDR}_{int}$ at the desired $q=0.2$ rate except for when $n=200$. DBL relies on asymptotic distributional approximations and can perform poorly when the sample size is small. Both K-Split and K-Full do a good job at controlling $\text{FDR}_{int}$ regardless of the sample size. For both estimators, $\text{FDR}_{int}$ is slightly above 0.2 when $n \in \{200,500\}$, though these estimators are intended to control a modified form of false discovery rates that is slightly less stringent, which leads to the performance here. HiGlasso, which does not control $\text{FDR}_{int}$ actually has $\text{FDR}_{int}$ close to, though slightly above, 0.2 in this scenario. Despite this, the power of HiGlasso to detect interactions is much lower than the proposed approaches. This shows that the proposed estimators provide a substantial increase in power without sacrificing in terms of increased false discovery rates. For Nlinteraction, the FDR is very close to 0 while the power is exactly 1 for all the 3 different sample sizes considered. 

\begin{figure}[htbp]
    \centering
    \includegraphics[width = 0.9\linewidth]{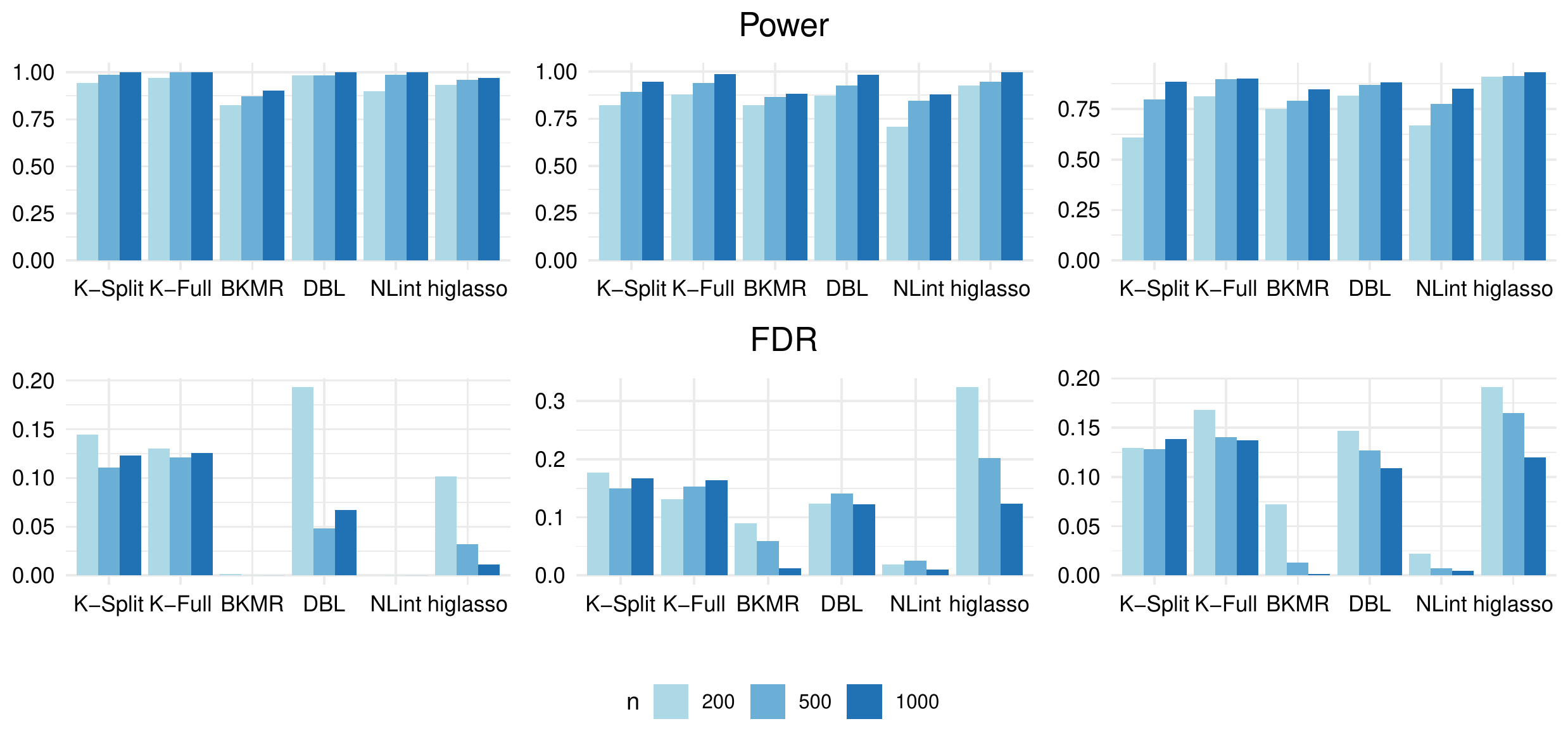}
    \caption{Overall FDR and power across all three simulation scenarios. The top panel displays the overall power, while the corresponding FDR rates are in the bottom panel. Each column corresponds to a particular simulation scenario. }
    \label{fig:overall}
\end{figure}

We now focus on traditional FDR and power for exposures having any association with the outcome. Figure \ref{fig:overall} shows that in all three scenarios, the proposed approaches have power that exceeds that of BKMR and Nlinteraction, and this does not come at a cost of overly high FDR, as our estimators have FDR below 0.2 in each simulation design. The proposed estimators have higher power than HiGlasso in scenario 1, but have slightly lower power for certain sample sizes under scenarios 2 and 3. We see, however, that HiGlasso also has FDR rates above the desired 0.2 level when $n=200$ in Scenario 2 showing the importance of explicitly controlling the FDR. 

\begin{figure}[htbp]
    \centering
    \includegraphics[width = 0.9\linewidth]{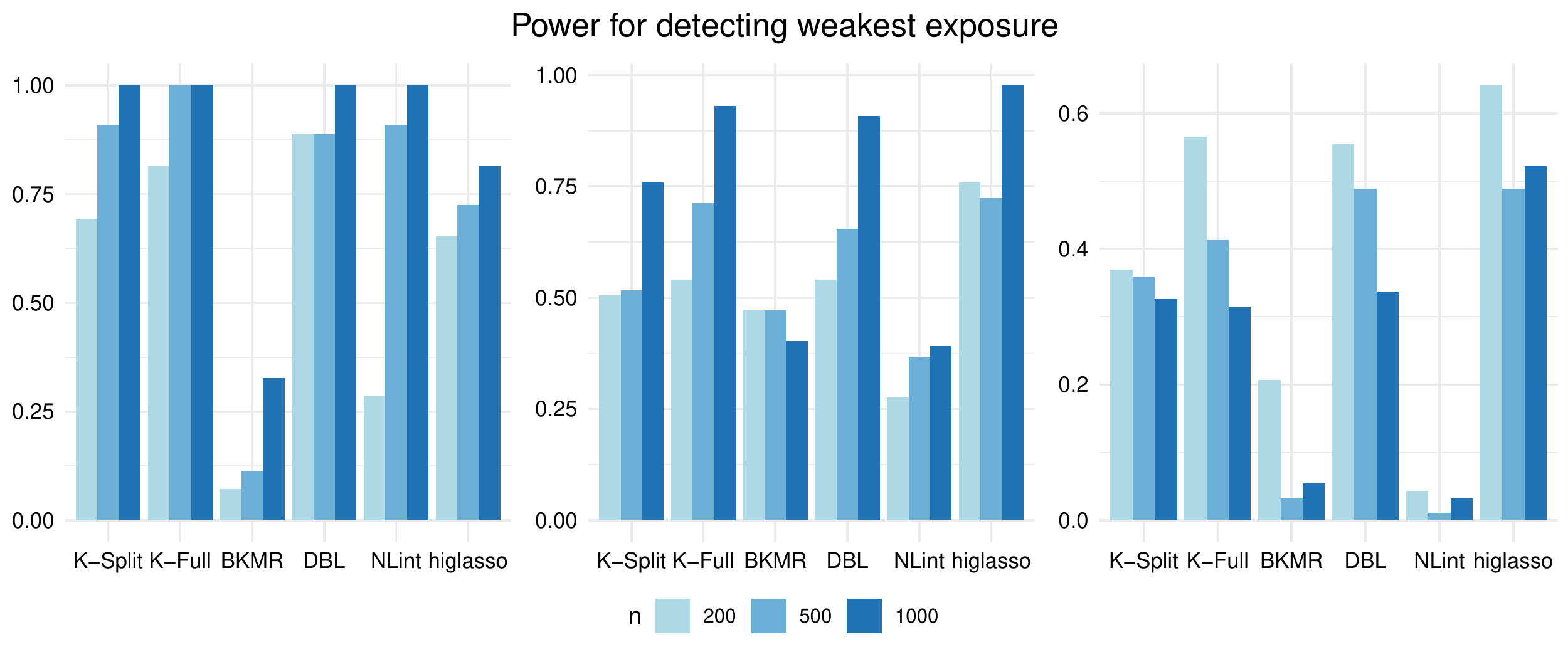}
    \caption{Power for detecting the exposure with the smallest signal. Each column corresponds to a particular simulation scenario.}
    \label{fig:power_weakest}
\end{figure}

A key feature of approaches is their ability to detect small signals, which are prevalent in environmental epidemiology. Figure \ref{fig:power_weakest} shows the power in each of the three scenarios to detect the exposure with the weakest signal-to-noise ratio. Regardless of the sample size, BKMR has very low power to detect small associations, and even as the sample size increases, BKMR does not necessarily have increasing power to detect these signals. Nlinteraction depicts a similar picture except for Scenario 1 where its power is comparable to that of the proposed methods for larger sample sizes. HiGlasso has moderate power in scenario 1 to detect the weakest exposure, but shows competitive performance to detect this exposure in scenarios 2 and 3. All of the proposed approaches, however, have substantial power to detect these associations regardless of the sample size or the scenario considered. This shows the importance of controlling the FDR when performing exposure selection. By controlling FDR, we are effectively optimizing power while maintaining a pre-specified rate of false discoveries.

\subsection{Mean squared error and coverage probabilities}

\begin{figure}[htbp]
    \centering
    \includegraphics[width = 0.8\linewidth]{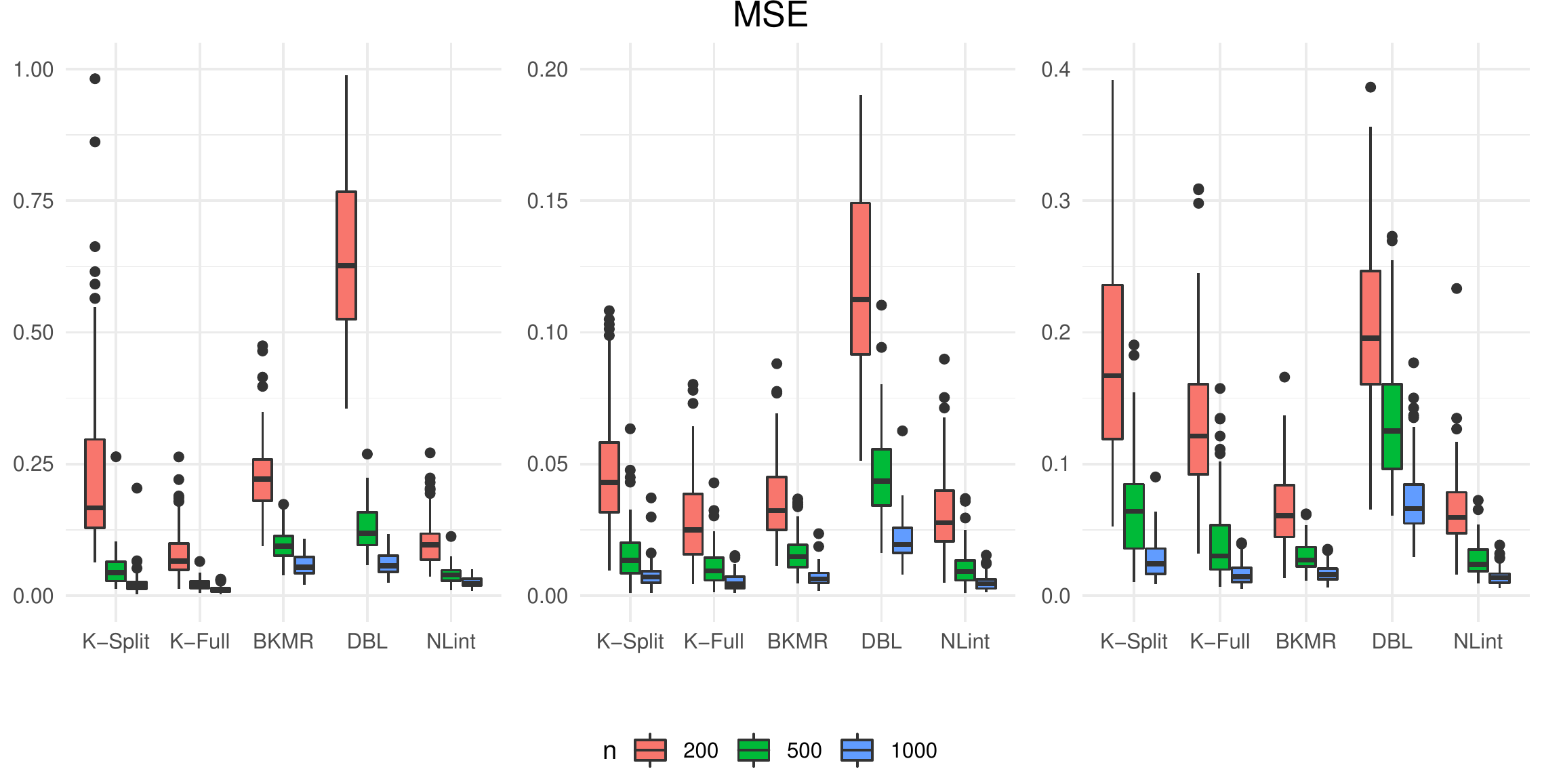} \\
    \includegraphics[width = 0.8\linewidth]{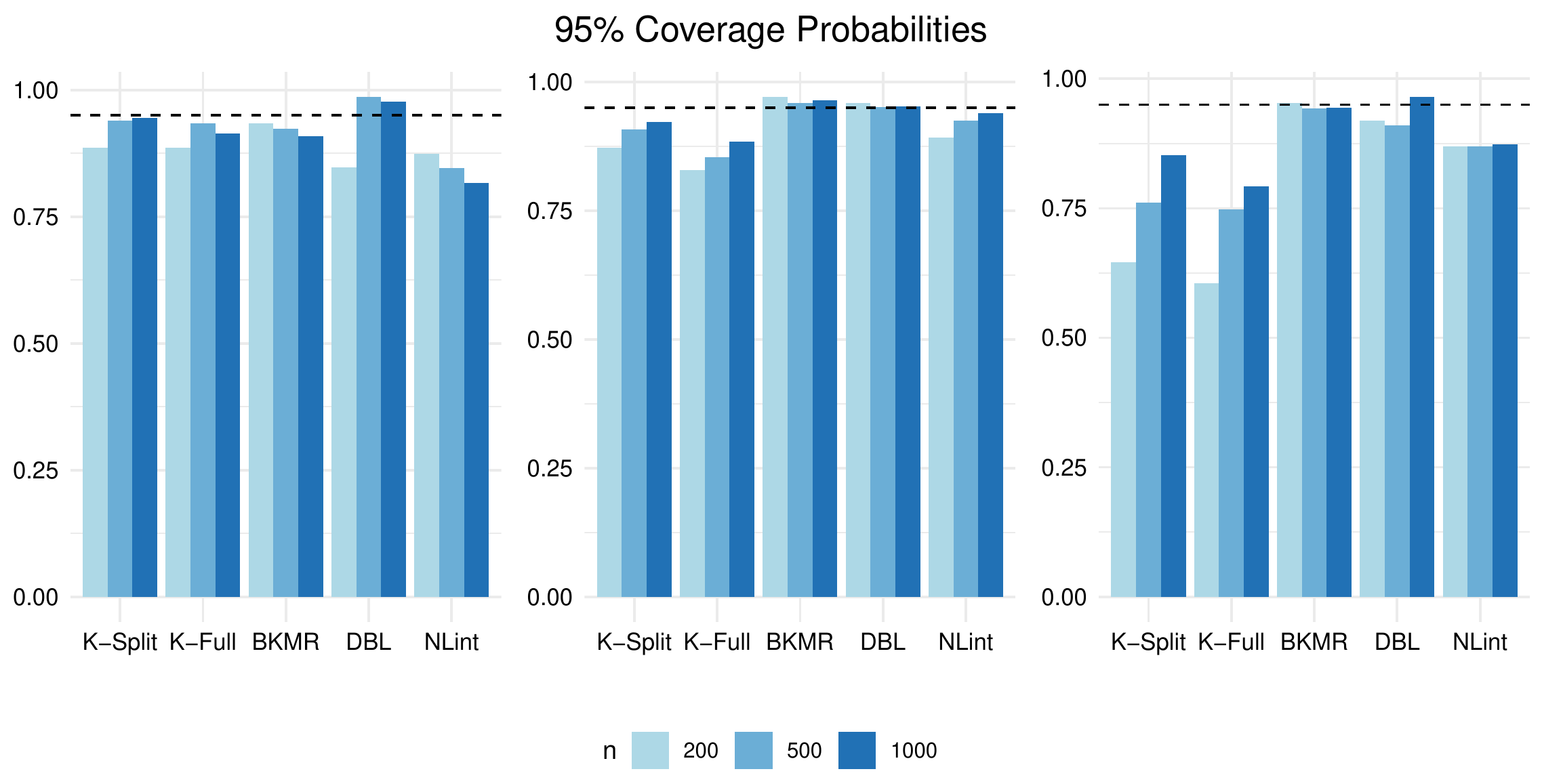}
\caption{MSE and 95\% coverage probabilities for estimating $f(\boldsymbol{X})$ across all three scenarios. The top panel corresponds to MSE, while the lower panel shows 95\% coverage probabilities. Each column corresponds to a particular simulation scenario.}
\label{fig:MSEcoverage}
\end{figure}

An equally important goal of mixtures analyses is to estimate the overall mixture effect, $f(\boldsymbol{X})$. Here, we focus on both MSE and 95\% interval coverages for $f(\boldsymbol{X}_i)$ averaged across all $n$ values of $\boldsymbol{X}_i$. Note that HiGlasso is not included in these results as it only identifies exposures and interactions and does not estimate the mixture effect. Figure \ref{fig:MSEcoverage} show boxplots of the MSE across simulations, as well as 95\% interval coverages.
BKMR performs relatively well in all scenarios considered, which is to be expected given that it is a flexible approach based on Gaussian processes that adapts well to different scenarios and has been shown to work well at estimating mixture effects. The DBL estimator suffers when the sample size is small ($n=200$), but improves greatly when $n=1000$. Despite this, it is outperformed by both BKMR and K-Full in all situations, regardless of sample size. K-Full performs quite well with respect to the MSE and slightly outperforms BKMR in scenarios 1 and 2. K-Split performs worse than K-Full in terms of MSE because it uses half of the sample size for both exposure selection and inference portions. Despite this, it still performs relatively well as long as the sample size is bigger than $n=200$, and even has lower MSE than BKMR in situation 1. When $n=200$, K-Split suffers as it only has 100 observations from which to perform exposure selection and inference, respectively, which negatively impacts estimation. 

In terms of interval coverage, all methods perform relatively well and obtain close to the nominal 95\% rate. Again DBL performs poorly when $n=200$, but this is to be expected as it relies heavily on asymptotic arguments, and does not perform well in small sample sizes. Both of the approaches based on knockoffs perform well with respect to 95\% interval coverage, except for scenario 3. The true function in scenario 3 contains sine and exponential functions that are not well approximated by the polynomial basis functions used in the proposed approaches, leading to worse performance. K-Full has slightly lower coverage than K-Split, especially in scenarios 2 and 3. K-Full uses the full data for both the exposure selection and subsequent inference on the chosen exposures. It is well known that ignoring the uncertainty from this first stage estimation can lead to anti-conservative inference, although there has been some recent work to suggest this can be ignored in certain settings \citep{zhao2017defense}. Nonetheless, our simulations here indicate that the anti-conservative nature of these intervals is not hugely problematic, and we still obtain coverage that is reasonably close to the desired level. 

\section{Analysis of NHANES data}

We utilize data from the 2001-2002 cycle of the National Health and Nutrition Examination Survey (NHANES), which is provided by the Centers for Disease Control and Prevention (CDC). NHANES is a nationally representative survey of the United States consisting of information from interviews, medical exams, and laboratory tests. We focus on a subset of the NHANES data examining the association of exposure to persistent organic pollutants and leukocyte telomere length (LTL) \citep{mitro2016cross}. This data measures exposure to polychlorinated biphenyls (PCBs), dioxins, and furans. PCBs were previously used as coolants or lubricants, but have been banned in the United States since the 1970s, though humans are still exposed through food \citep{faroon2000toxicological}. Both dioxins and furans are introduced as byproducts when other chemicals are produced. Dioxins and PCBs are both carcinogenic, though the mechanisms by which exposure to these chemicals leads to increased cancer risk is not fully understood \citep{faroon2000toxicological, international2012chemical, mitro2016cross}. Telomeres are protective segments of DNA at the ends of chromosomes, and their lengths generally decrease with age. While telomere length typically decreases, they can be elongated, and environmental pollutants have been associated with both increases and decreases in telomere length. Telomerase, which can increase telomere length, is generally low in normal cells, but is more active in cancer cells. This shows the importance of understanding how these exposures and LTL are associated, as this could be related to cancer risk. 

Our data was previously analyzed in \cite{gibson2019overview}, which reviewed the statistical approaches for environmental mixtures, and used them to analyze the associations between persistent organic pollutants and LTL. They identified a number of exposure effects, most of which were linear with increasing levels of environmental exposures being associated with increases in LTL. They did not find evidence of interactions among the exposures, though the only approach considered that allows for interactions was BKMR, which does not identify important interactions and can only provide visual evidence of the presence of an interaction. Additionally, none of the approaches considered false discovery rates when identifying important exposures. We will apply our approaches to identifying both important exposures and interactions and compare results to the existing approaches that do not target FDR rates or interactions. 

Our data contains 1003 subjects with measurements of 18 environmental exposures, demographic information, and the results of laboratory tests on blood counts. Specifically, we have measurements on 11 PCBs, 3 dioxins, and 4 furans. Additionally, we will adjust for age, BMI, education, sex, race, lymphocyte count, basophil count, eosinophil count, cotinine level, monocyte count, and neutrophil count. Our goal will be to assess whether any of the 18 environmental exposures are conditionally associated with LTL, and whether or not these associations include interactions among the pollutants. We apply BKMR, NLint, and HiGlasso as well as the DBL, K-Split, and K-full approaches with $q$ set to 0.2. 

\subsection{Exposure selection results}

\begin{figure}[htbp]
    \centering
    \includegraphics[width = 0.8\linewidth]{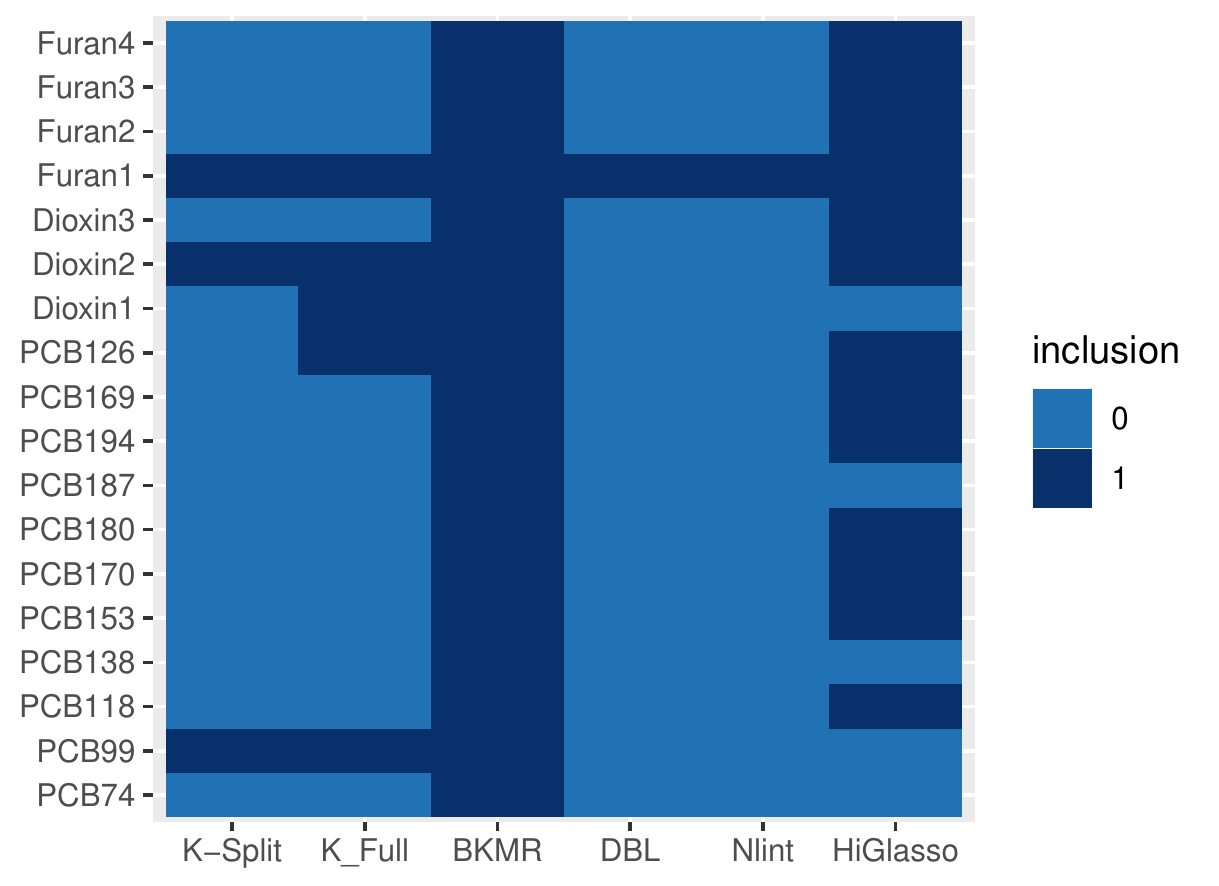} 
\caption{Inclusion of exposures into the respective models for the NHANES analysis.}
\label{fig:varSelection}
\end{figure}

The inclusion and exclusion of exposures in the model across all methods considered can be found in Figure \ref{fig:varSelection}. One interesting result is that BMKR selects every exposure to be included in this analysis. BKMR selects an exposure if the posterior inclusion probability for that exposure is greater than 0.5, and every posterior inclusion probability was above 0.84 here. We considered the default prior distribution when using BKMR, though results are highly sensitive to this prior specification, and would likely change if we altered the prior distribution. Not shown are the estimated exposure response curves for each exposure from the BKMR model fit, but nearly all of them show a flat, linear trend. This indicates that the mixture effect is being overly shrunk towards zero. HiGlasso also selects a large number of exposures as 13 of the 18 exposures are deemed significant. The other approaches lead to more sparse solutions as NLint only selects one exposure, while the proposed approaches select 3 (K-split), 5 (K-full), and 1 (DBL). All of the approaches select furan 1, indicating a clear association between this pollutant and telomere length. 

In terms of interactions, both HiGlasso and NLint choose zero interactions. Neither of these approaches target FDR control, and as we have seen in the simulation, they have very low FDR rates for interactions, which can result in a reduction in power. DBL also does not include any interactions into the model, however, the K-split approach, which has lower power due to the splitting of the  data before identifying interactions, identifies the interaction between PCB 99 and dioxin 2. The K-Full approach also identifies this interaction, and the interaction between dioxin 1 and dioxin 2. The knockoffs approach controls the false discovery rate regardless of correct model specification, which gives us increased confidence that these identified exposures are real signals that would not otherwise be identified by existing approaches. 

\subsection{Estimated exposure-response surfaces}

\begin{figure}[htbp]
    \centering
    \includegraphics[width = 0.45\linewidth]{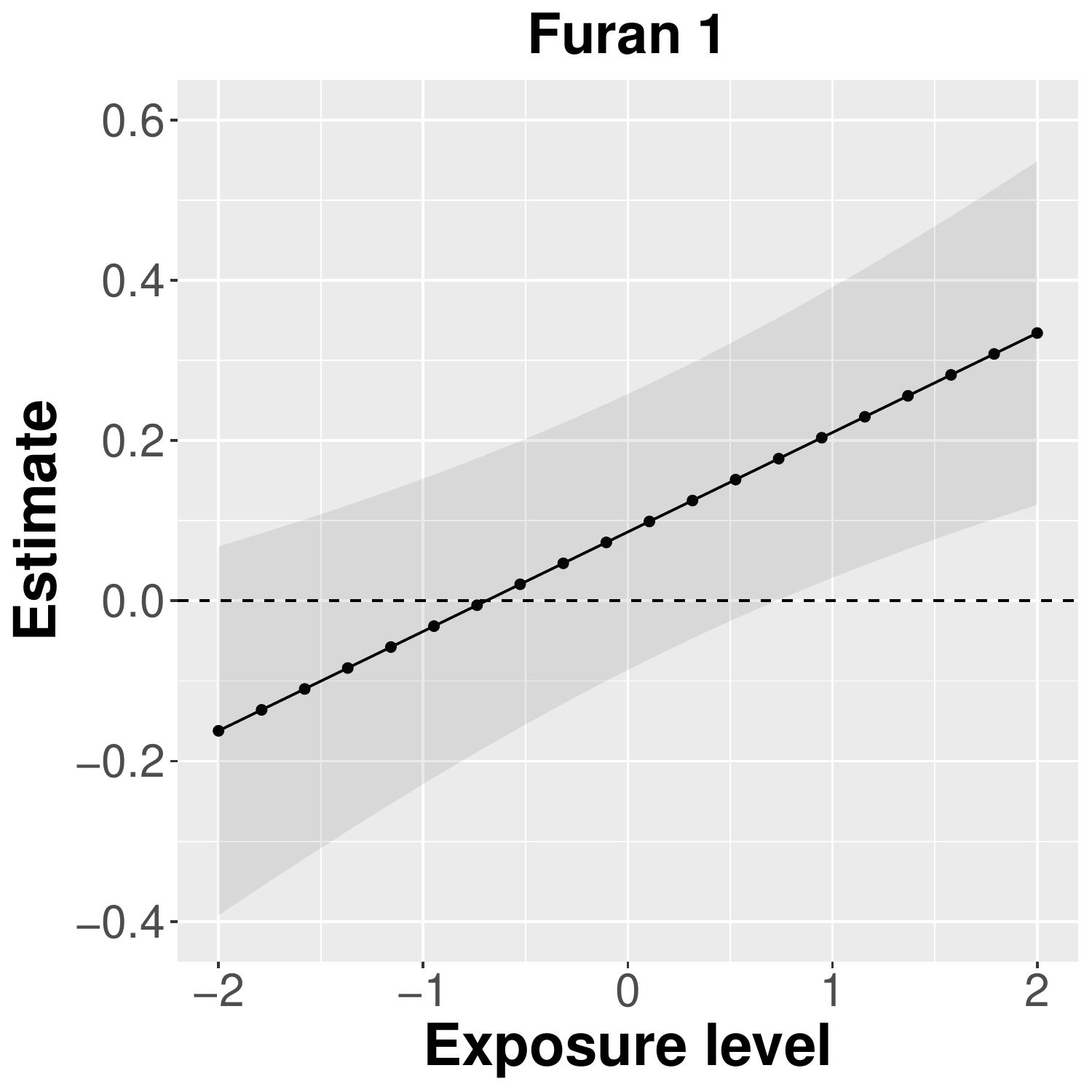}
    \includegraphics[width = 0.45\linewidth]{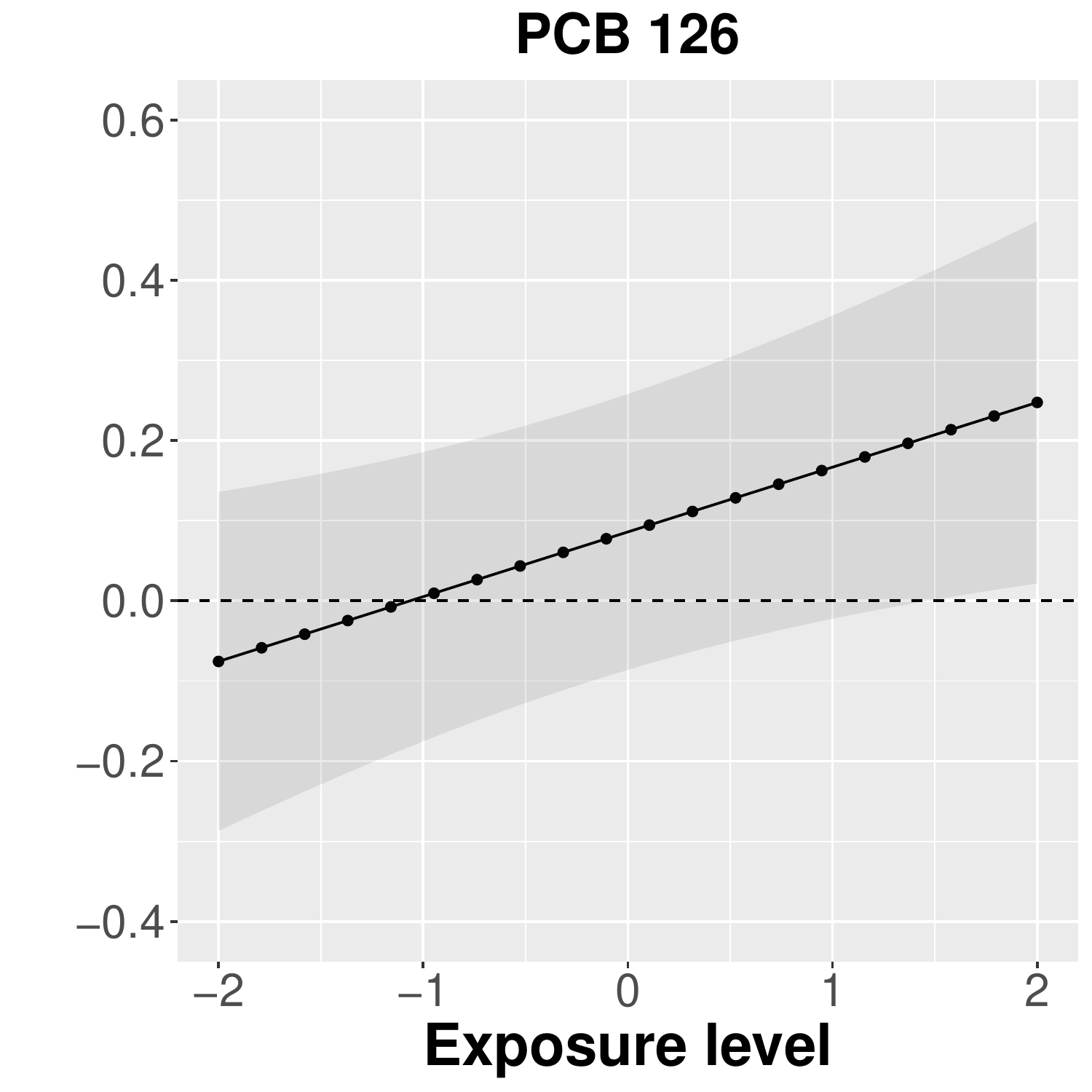}
\caption{Plot showing the marginal effect of exposure to both Furan 1 and PCB 126 on average telomere length along with pointwise 95\% confidence intervals for the K-Full estimator.}
\label{fig:NHANESmainEffect}
\end{figure}

Now we shift focus to the estimation of the effects of the exposures identified as important. In particular, we will examine exposure response curves for the individual exposures, as well as exposure response surfaces between any two exposures for which an interaction was identified. To estimate exposure response curves for a single exposure, we will fix the values of the other exposures at their mean value, and vary the value of the exposure of interest to see how $f(\boldsymbol{X})$ changes as a function of this exposure. We perform the same procedure for interactions, though we vary the values of both exposures on a two-dimensional grid, while holding constant the values of the remaining exposures and examine $f(\boldsymbol{X})$. We focus on the exposures and interactions identified by the K-Full procedure as this was the best performing approach in simulations among the three that we proposed in this paper. K-full identifies both PCB 126 and furan 1 as important exposures that are not included as interaction terms, and the exposure response curves for both of these can be found in Figure \ref{fig:NHANESmainEffect}. For both pollutants, we see an increasing, linear trend suggesting that higher levels of both these pollutants are associated with increases in telomere length. Note, the x-axis ranges from -2 to 2 as we are dealing with centered and scaled exposure values.

\begin{figure}[htbp]
    \centering
    \includegraphics[width = 0.9\linewidth]{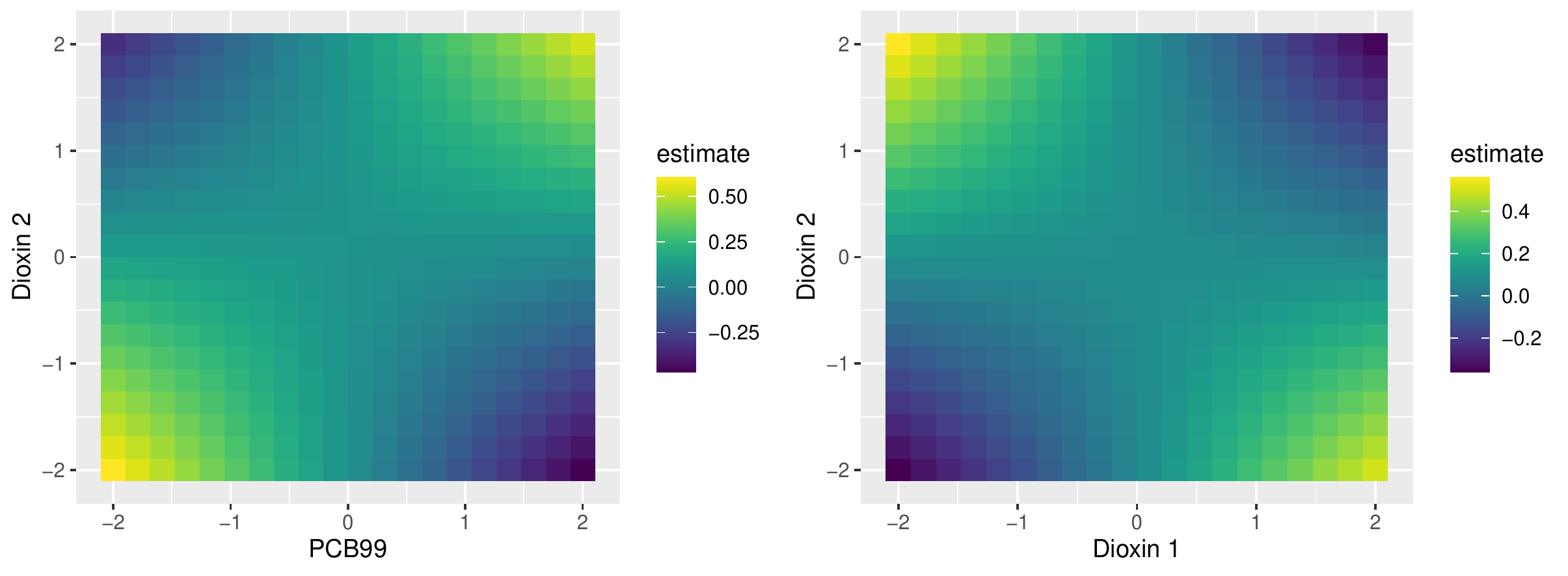} 
\caption{Plots showing the interaction effect between PCB 99 and Dioxin 2 (left) and between Dioxin 1 and Dioxin 2 (right) on the average telomere length for the K-Full estimator.}
\label{fig:intKF}
\end{figure}

As for the interactions, K-Full selects interactions between PCB 99 and dioxin 2 and between dioxin 1 and dioxin 2. The exposure response surfaces for both of these pairs can be found in Figure \ref{fig:intKF}. Looking at the left panel, one can see that $f(\boldsymbol{X})$ is largest when both PCB 99 and dioxin 2 have similar values, and attains its peak when both exposures are either very high or very low in the range of values seen in our data. We see a somewhat different interaction effect for dioxin 1 and dioxin 2, which indicates an inverse relationship between the two exposures. When all other exposures are set to their mean values, $f(\boldsymbol{X})$ is largest when either dioxin 1 is high and dioxin 2 is low, or vice-versa. 

\section{Discussion}

We have described two approaches to estimating the effects of environmental mixtures in a way that identifies important exposures and interactions while controlling false discovery rates. By utilizing recently developed approaches in the high-dimensional statistics literature, we can simultaneously identify important exposures and perform inference on the mixture effect. Many approaches in the literature either do not perform selection on interactions (BKMR), or do not perform inference on the mixture effect (HiGlasso). While there are approaches that do both \citep{herring2010nonparametric,antonelli2020estimating}, none of these approaches control false discovery rates. We have seen that controlling false discovery rates can greatly improve power, particularly to identify the effects of environmental exposures that have small associations with the outcome, which are common in environmental epidemiology. Additionally, our approaches are fully automated and do not require selection of hyperprior parameters that can influence the resulting mixture effect or set of exposures identified as important. 

Our approaches come with certain limitations. The debiased lasso makes strong assumptions about the underlying sparsity in the data generating model and requires large sample sizes. Additionally, it does not provide as precise of estimates of the overall mixture effect when compared to BKMR or the knockoffs approaches considered here. For this reason, we recommend using the K-Full or K-Split approaches which perform well in terms of power, false discovery rates, MSE, and coverage probabilities. K-Full tends to outperform K-Split as it uses the full sample for both exposure/interaction selection and inference on the subsequent exposure effect conditional on the chosen exposures. The main drawback of K-Full is that the confidence intervals on the effect of the environmental exposure are not theoretically valid as the data is being used in both the exposure selection and post-selection inference steps. This issue can be avoided by using the K-Split or DBL approaches, however, these will either have reduced power to detect effects (K-Split), or require larger sample sizes before giving reasonable estimates (DBL). Empirically, we have seen that K-Full gives reasonable assessments of uncertainty that give confidence interval coverage close to the desired rate, and it has the best power and MSE of the approaches considered. 

\bibliographystyle{apalike}
\bibliography{References}

\appendix

\section{Additional simulation results}

In Section 4 we highlighted the performance of the proposed methods (K-Split, K-Full and DBL) along with existing approaches in the literature across different metrics. There we considered three data generating mechanisms and presented results for $p=10$. Here, in Section \ref{subsec:p20} we present simulation results from the same scenarios but with $p=20$. We additionally consider sample sizes, $n \in \{200,500,850,1000\}$. The overall pattern in the results remains unchanged from the $p=10$ situation with a few minor differences. We also present the simulation results from an additional data generating process in Section \ref{sec:addsim}. 

\subsection{Simulation Results for $p=20$}
\label{subsec:p20}

\begin{figure}[htbp]
    \centering
    \includegraphics[scale = 0.5]{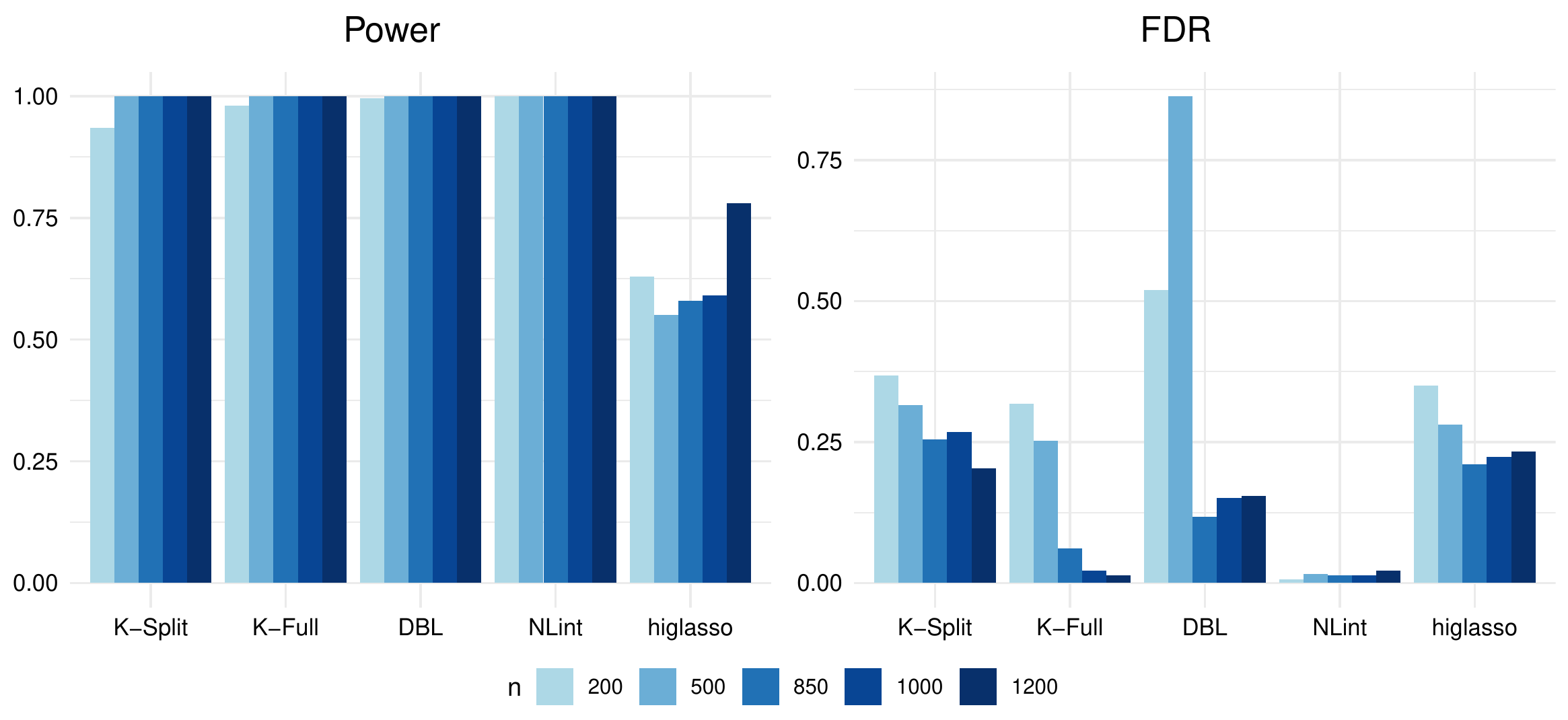}
    \caption{Empirical Power and FDR for interactions for simulation scenario 1 in the manuscript with $p = 20$. The left panel corresponds to power to detect interactions while the right panel shows $\text{FDR}_{int}$.}
    \label{fig:int20}
\end{figure}

Figure \ref{fig:int20} shows the power and false discovery rates for detecting important interaction pairs from simulation scenario 1. For K-Split and K-Full, $\text{FDR}_{int}$  is around the desired level of $q=0.2$ though it is slightly larger in certain settings likely due to the fact that we are controlling the modified false discovery rate. DBL also controls $\text{FDR}_{int}$ effectively at $q=0.2$ except for small sample sizes as it heavily relies on asymptotic distributional approximations. HiGlasso has similar $\text{FDR}_{int}$ to that of K-Split but the power of HiGlasso to detect interactions is much lower than the proposed approaches. For Nlinteraction, once again the FDR is very close to 0 and the power is 1 for all the different sample sizes considered.

\begin{figure}[htbp]
    \centering
    \includegraphics[scale = 0.5]{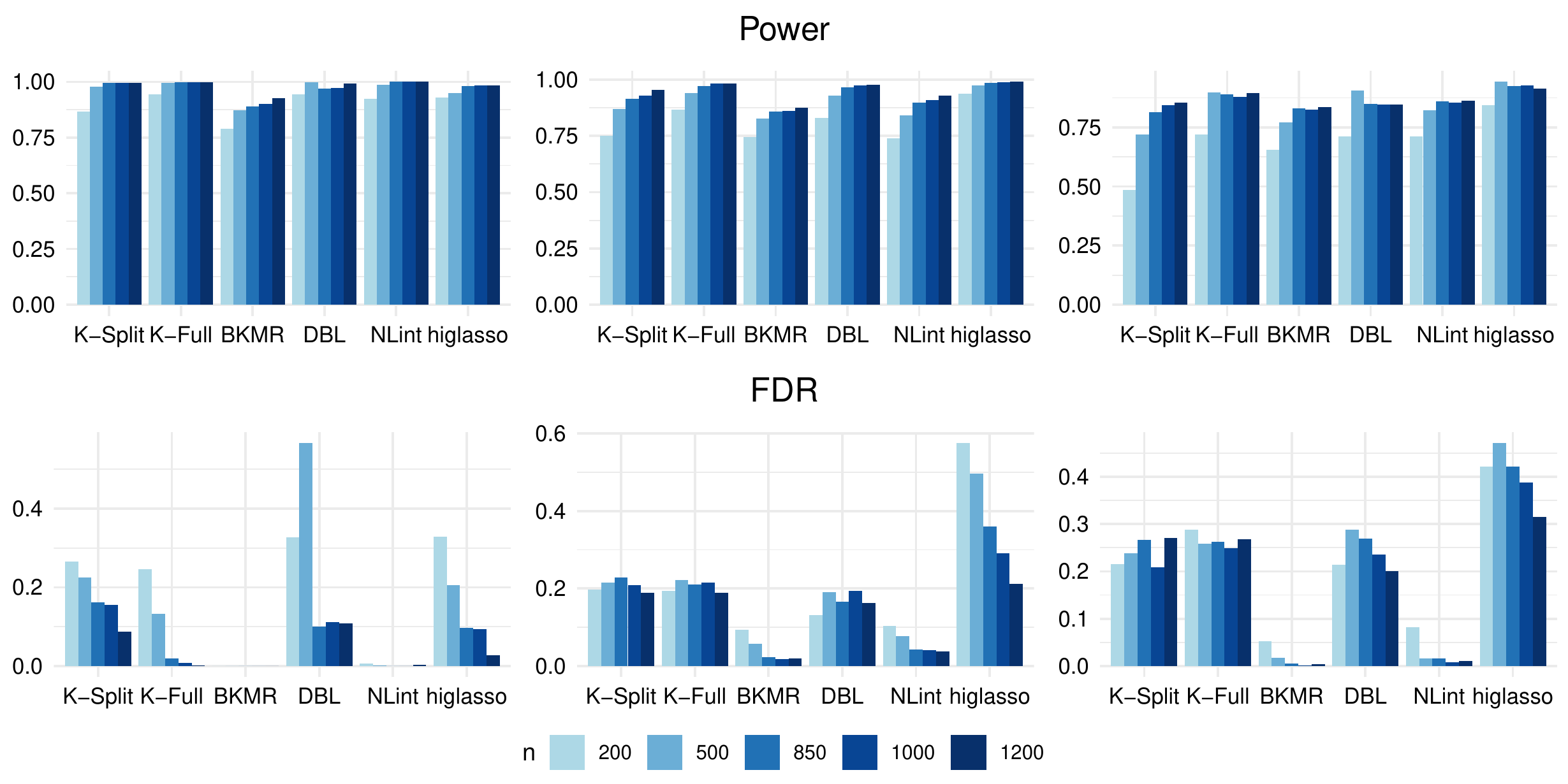}
    \caption{Empirical overall Power and FDR for settings 1,2 and 3 with $p = 20$; Overall Power on the first row and the corresponding overall FDR on the second row.}
    \label{fig:overall20}
\end{figure}

 Figure \ref{fig:overall20} shows the traditional FDR and power for each of the three scenarios. In all scenarios, BKMR is outperformed by the proposed approaches in terms of power, and the FDR for these methods are below 0.2 except for scenario 3 where the FDR is slightly above 0.2. NLinteraction displays competitive performance but in scenario 2 the proposed methods have better power. HiGlasso is outperformed by the proposed methods in terms of power for scenario 1, while it performs better in the other two scenarios. This, however, is mostly because it has high FDR values caused by the fact that it does not attempt to control FDR.
 
\begin{figure}[htbp]
    \centering
    \includegraphics[scale = 0.55]{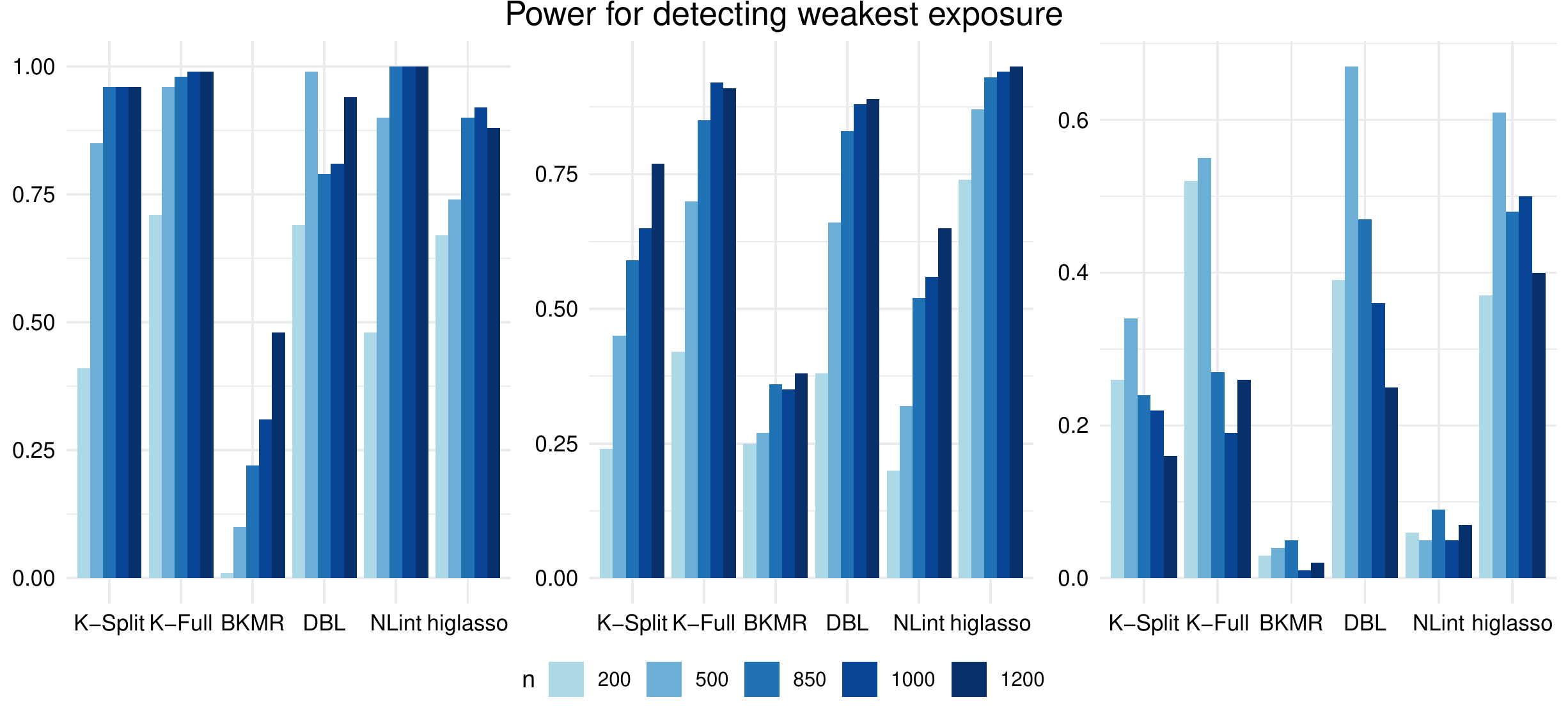}
    \caption{Power for detecting the weakest exposure for settings 1,2 and 3 with $p = 20$.}
    \label{fig:weakest20}
\end{figure}

 Figure \ref{fig:weakest20} shows the power to detect the exposure that has the weakest association with the outcome in each of the three scenarios. BKMR displays low power to detect these small to moderate associations in all three scenarios regardless of sample size. Even though the power increases with sample size for BKMR in the first two scenarios it is not quite as high as the other approaches. Nlinteraction performs well in Scenario 1 but suffers in the other two scenarios. HiGlasso also shows competitive performance to detect the weakest exposure in all the scenarios, but also has inflated FDR leading to artificially high power rates. Once again this exhibits how controlling FDR can greatly impact the power to detect small to moderate exposure effects.

\begin{figure}[htbp]
    \centering
    \includegraphics[scale = 0.5]{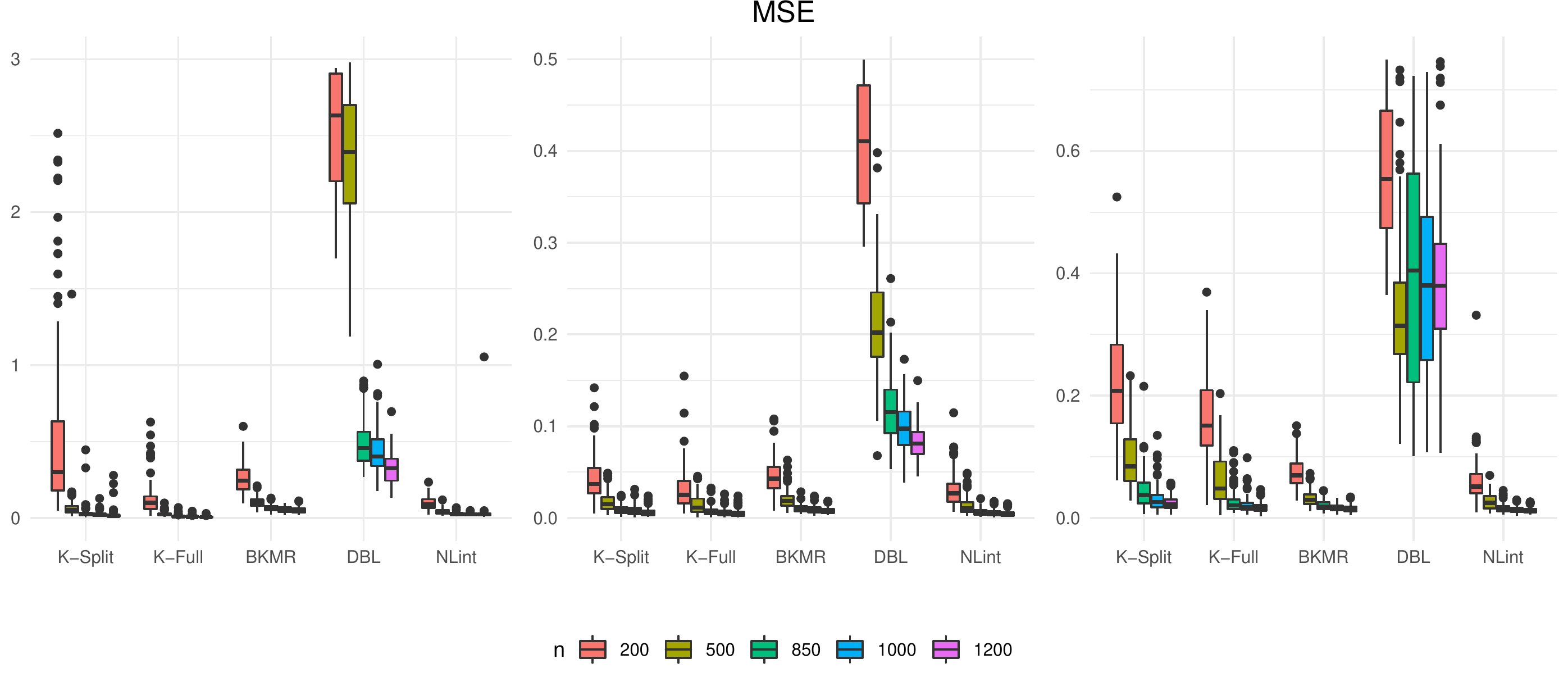}
    \caption{MSE for settings 1,2 and 3 with $p = 20$.}
    \label{fig:MSE20}
\end{figure}

 Figure \ref{fig:MSE20} shows boxplots of the MSE from each simulated data set across each scenario and sample size. Once again HiGlasso is not included in these simulations as it does not estimate or perform inference on the resulting mixture effect. BKMR being a flexible approach based on Gaussian processes performs relatively well in all scenarios considered, as was the case for the $p=10$ situation. Again the DBL estimator suffers for smaller sample sizes and takes larger samples for the MSE to be stable. K-Full performs quite well with respect to the MSE for $f(\boldsymbol{X})$ and it also slightly outperforms BKMR in scenarios 1 and 2, while K-Split performs worse than K-Full in terms of MSE because it uses half the sample size for both the exposure selection and inference portions. These results closely mirror those seen in the main manuscript with $p=10$.

\begin{figure}[htbp]
    \centering
    \includegraphics[scale = 0.55]{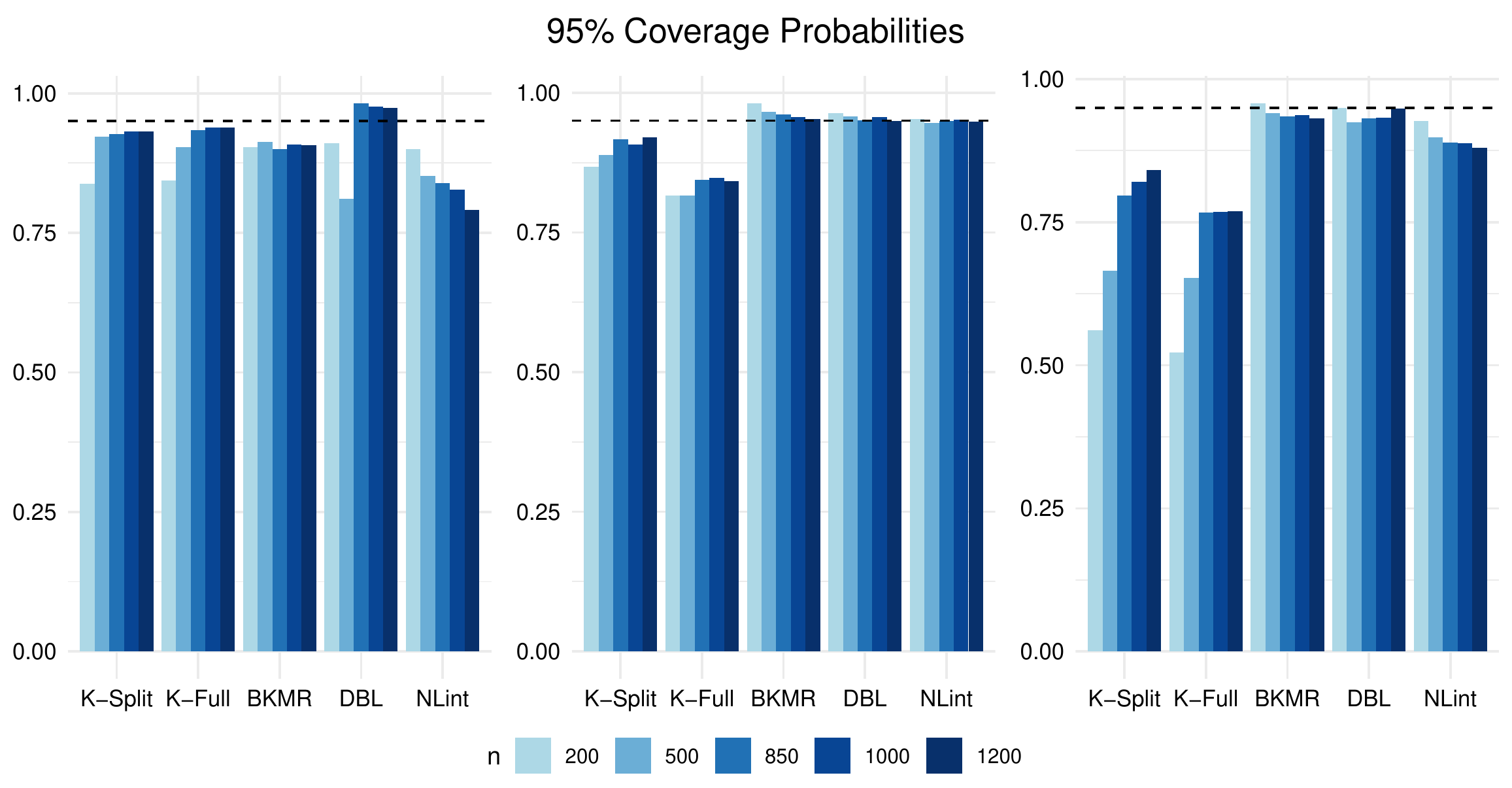}
    \caption{95\% coverage for settings 1,2 and 3 with $p = 20$.}
    \label{fig:coverage20}
\end{figure}

Figure \ref{fig:coverage20} shows the 95\% interval coverage for the 3 scenarios and the five different sample sizes. All five methods perform relatively well by obtaining close to, or above the nominal 95\% rate. The pattern is similar to what we saw in the main manuscript with $p=10$. DBL performs poorly for small sample sizes in scenario 1. K-Split and K-Full perform relatively well except for scenario 3.

\subsection{Additional Simulation Scenario}
\label{sec:addsim}

\begin{figure}[htbp]
    \centering
    \includegraphics[scale = 0.7]{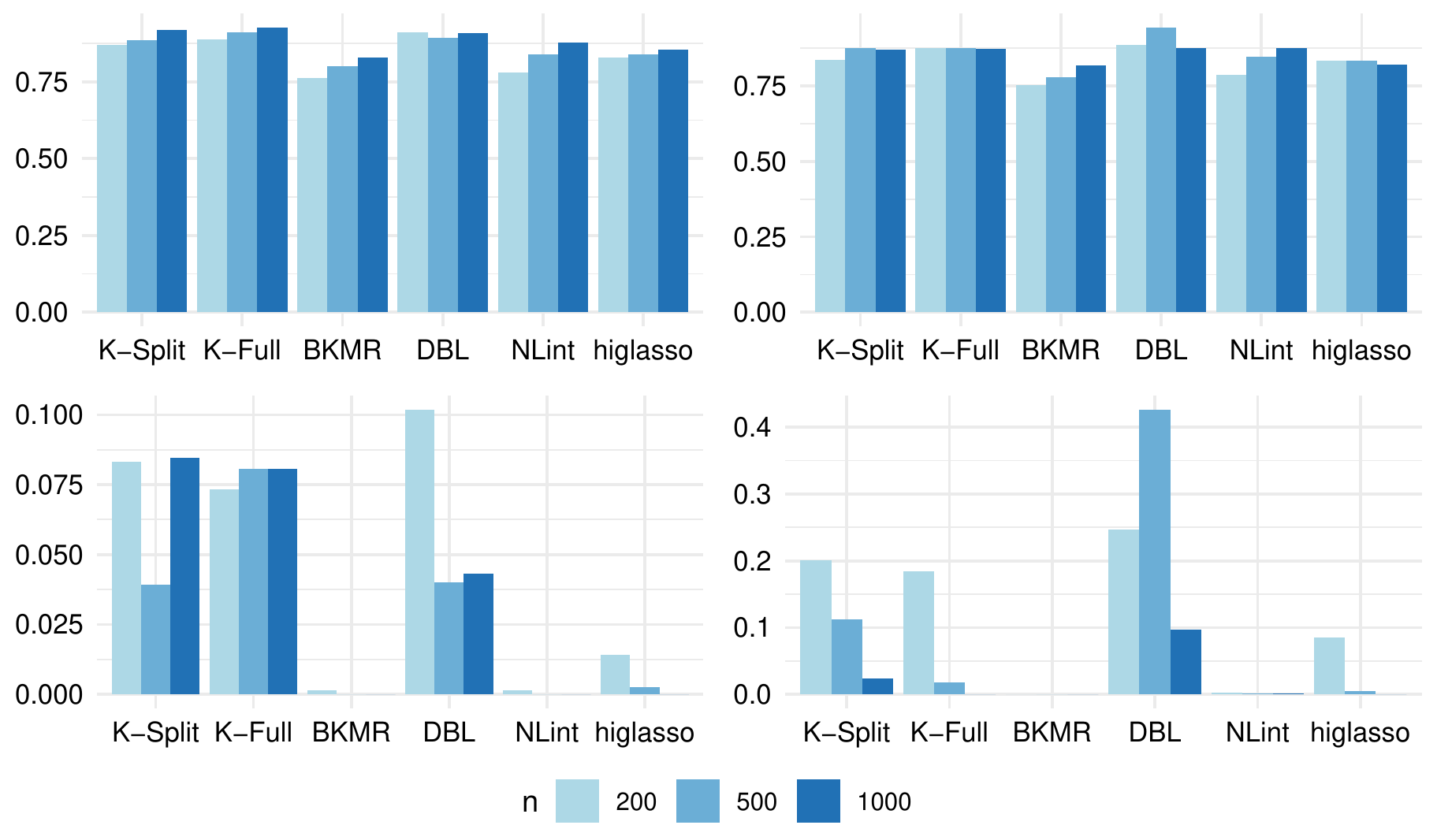}
    \caption{Empirical overall Power and FDR for the additional setting with $p = 10$ and $p = 20$; Overall Power is on the first row and the corresponding FDR is on the second row. The first column corresponds to $p=10$, while the second column corresponds to $p=20$.}
    \label{fig:overallapp}
\end{figure}

Here we consider the following true function: $ f(\mathbf{x}) = 2.3x_1 + 1.9x_1^2 - 0.2x_2 + 0.4x_2^2  - 1.5x_4 - 1.5x_4^2 + 0.05x_5 + 1.2x_7 + 0.8x_8 + x_9 + 0.2x_{10} + 0.1x_{10}^2$ with $n \in \{200,500,1000\}$ and $p \in \{10,20\}$. In Figure \ref{fig:overallapp} we present the traditional FDR and power for $p=10$ and $p=20$. We see the proposed methods are performing better than BKMR in terms of overall power for both situations. K-Split and K-Full are able to control the FDR at the desired level 0f 0.2. DBL is also able to control FDR except for smaller sample sizes ($n=500$) when $p=20$. HiGlasso and NLint show competitive performance for both $p=10$ and $p=20$, though their power is less than for the proposed approaches.

\begin{figure}[htbp]
    \centering
    \includegraphics[scale = 0.6]{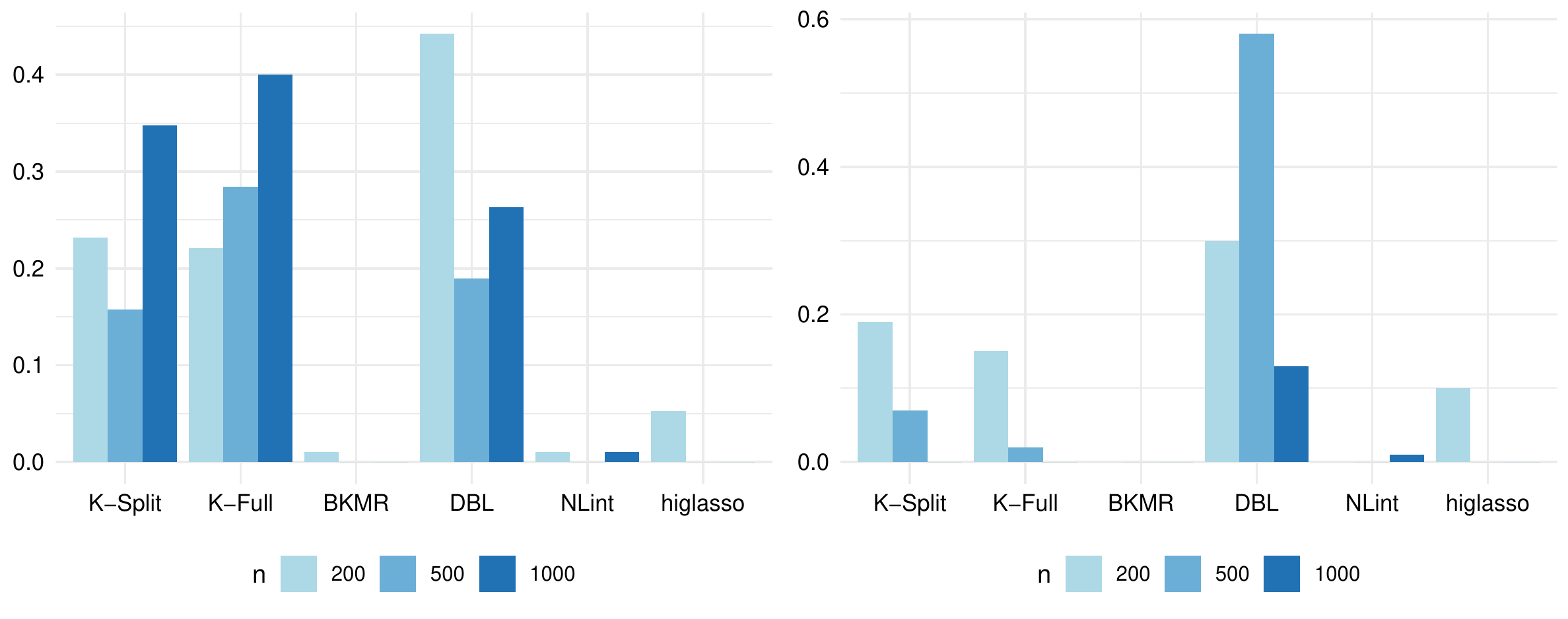}
    \caption{Power for detecting the weakest exposure for the additional setting with $p = 10$ (Left) and $p = 20$ (Right).}
    \label{fig:weakestapp}
\end{figure}

Figure \ref{fig:weakestapp} shows the power for detecting the weakest exposure for both $p=10$ and $p=20$. We see for $p=10$ the proposed methods are significantly better than the other approaches. For $p=20$ most of the methods are unable to detect the weakest exposure, though the proposed methods do still maintain the largest power.  

\begin{figure}[htbp]
    \centering
    \includegraphics[scale = 0.6]{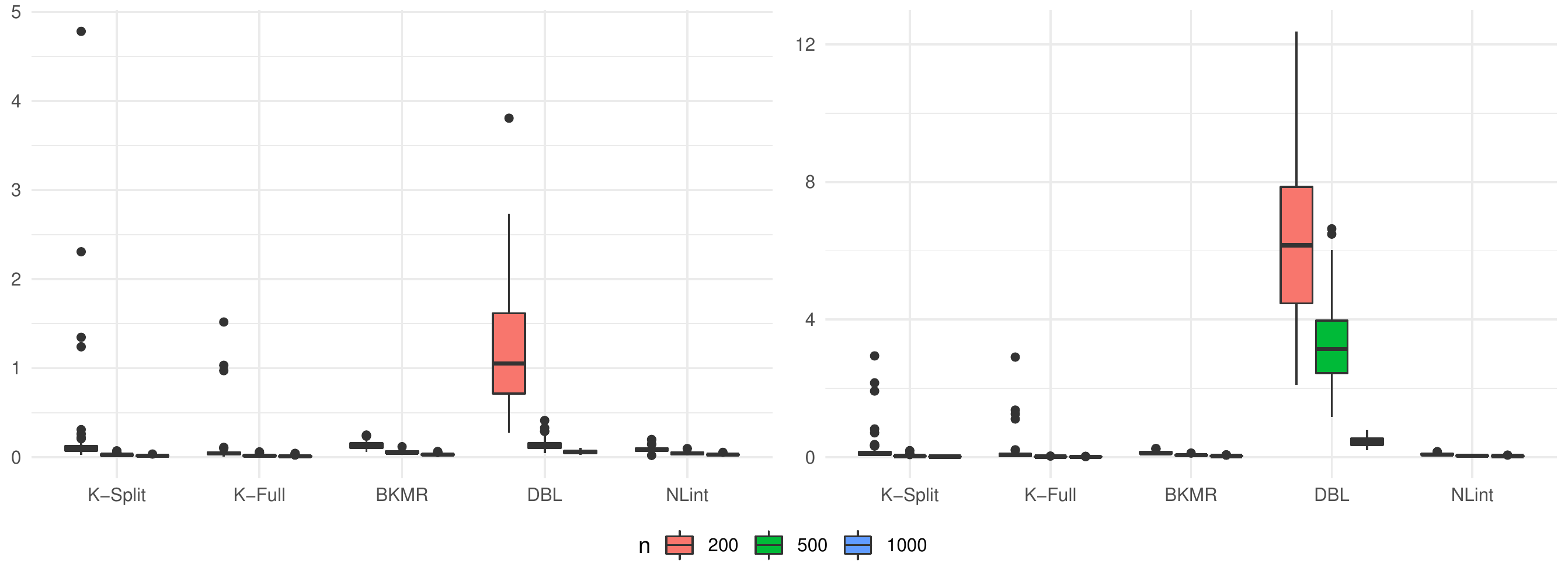}
    \caption{MSE for estimating $f(\boldsymbol{X})$ in the additional setting with $p = 10$ (Left) and $p = 20$ (Right).}
    \label{fig:MSEapp}
\end{figure}

Figure \ref{fig:MSEapp} shows boxplots of MSE for estimating  $f(\boldsymbol{X})$. K-Full has the lowest MSE for both $p=10$ (Left) and $p=20$ (Right). DBL exhibits larger MSE for smaller sample sizes and improves when $n$ reaches 1000, just as was the case for the other scenarios considered.

\begin{figure}[htbp]
    \centering
    \includegraphics[scale = 0.7]{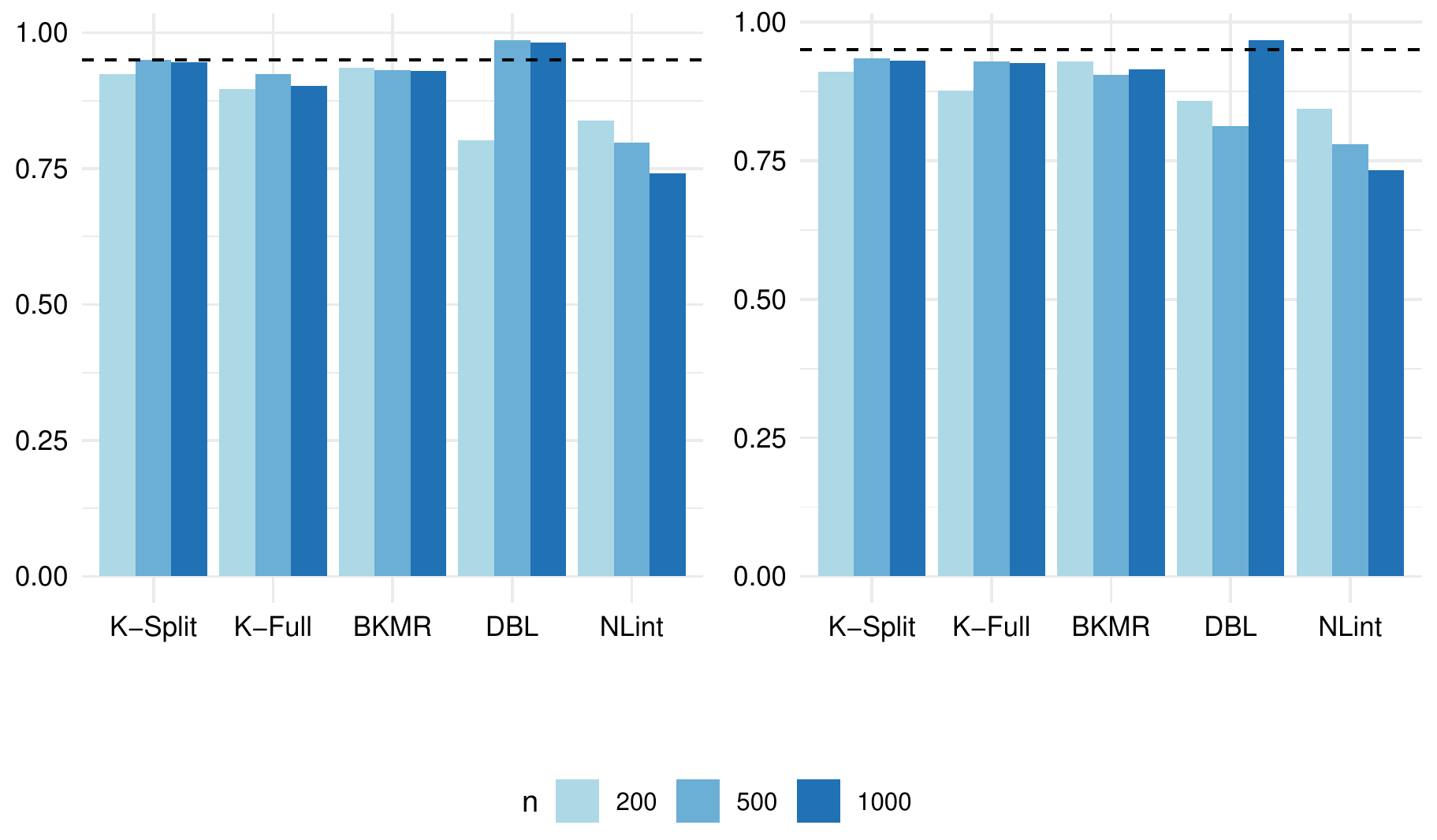}
    \caption{95\% coverage for estimating $f(\boldsymbol{X})$ in the additional setting with $p = 10$ (Left) and $p = 20$ (Right).}
    \label{fig:coverageapp}
\end{figure}

Figure \ref{fig:coverageapp} shows 95\% coverage for estimating  $f(\boldsymbol{X})$ for all the methods when $p=10$ and $p=20$. As was seen in the other scenarios, DBL does worse for smaller $n$ values but improves as the sample size increases. K-Split performs better than K-Full for reasons described in Section 3.3. NLint has lower coverages than all of the other methods here. Overall, all of the proposed methods attain the desired 95\% coverage for both $p=10$ and $p=20$.

\section{Derivation for upper bound on the false discovery rate}

Here we present the steps leading to the bound on FDP given in Section 3.4 of the manuscript. Let FD and D be the total number of overall (not interaction) false discoveries and discoveries, respectively. Let $O_d$ and $O_n$ be defined as in Section 3.4. Here we restrict our attention to the selected interactions only and corresponding overall exposures that are chosen due to their inclusion into at least one interaction. For this derivation we focus on these selected exposures and ignore any exposures that are selected from the main effects component of the model. Then,
\begin{align*}
    \text{FDP}_{int} &= \frac{\text{\# false interaction discoveries}}{\text{\# interaction discoveries}} \\
    &= \frac{\frac{1}{2}\text{FD} + O_n}{\frac{1}{2}\text{D} + O_d}\\
     &\geq \frac{\frac{1}{2}\text{FD}}{\frac{1}{2}\text{D}+ O_d}\\
     \Rightarrow \text{FDP}_{int} \bigg( \frac{1}{2}\text{D}+ O_d \bigg) &\geq \frac{1}{2}\text{FD} \\
     \Rightarrow \text{FDP}_{int} \bigg( \frac{\frac{1}{2}\text{D}+ O_d}{\frac{1}{2} D} \bigg) &\geq \frac{\frac{1}{2}\text{FD}}{\frac{1}{2}\text{D}} \\
     \Rightarrow \text{FDP}_{int} \bigg( 1 + \frac{2 O_d}{D} \bigg) &\geq \text{FDP} \\
\end{align*}
and we have the desired result. Once we have this result, we can use a straightforward application of the Cauchy-Schwartz inequality to see that

\begin{align*}
    \text{FDR} = E(\text{FDP}) &\leq E \Bigg( \text{FDP}_{int} \bigg( 1 + \frac{2 O_d}{D} \bigg) \Bigg) \\
    &\leq \sqrt{E(FDP_{int}^2) E \Bigg( \Big[ 1 + \frac{2 O_d}{\text{D}} \Big]^2 \Bigg)} \\
\end{align*}

\end{document}